# Agent-based Computational Economics in Management Accounting Research: Opportunities and Difficulties

Friederike Wall✉ and Stephan Leitner
University of Klagenfurt
Department of Management Control and Strategic Management

**ABSTRACT**

Agent-based computational economics (ACE) – while adopted comparably widely in other domains of managerial science – is a rather novel paradigm for management accounting research (MAR). This paper provides an overview of opportunities and difficulties that ACE may have for research in management accounting and, in particular, introduces a framework that researchers in management accounting may employ when considering ACE as a paradigm for their particular research endeavor. The framework builds on the two interrelated paradigmatic elements of ACE: a set of theoretical assumptions on economic agents and the approach of agent-based modeling. Particular focus is put on contrasting opportunities and difficulties of ACE in comparison to other research methods employed in MAR.

**Acknowledgement**: The authors thank the Senior Editor, Eva Labro, and three anonymous reviewers for their valuable comments and suggestions.

---

✉ corresponding author: friederike.wall@aau.at

# INTRODUCTION

Theoretical foundations and methods adopted in management accounting research (MAR) are manifold. According to studies on MAR (e.g., Shields 1997, Hesford et al. 2007, Lindquist and Smith 2009, Hopper and Bui 2016, Guffey and Harp 2017), theoretical foundations range from economics, psychology and cognitive sciences to sociology, and the methods employed comprise, for example, analytical, survey-based, experimental, conceptual, case-based and archival approaches.

Even though economics has witnessed the advent of Agent-Based Computational Economics (ACE) and ACE has been applied extensively to various domains of managerial science, the application of this perspective to MAR has remained limited.[1] This paper provides an overview of opportunities and difficulties that ACE may have for research in management accounting and, in particular, introduces a framework that researchers in management accounting may employ when considering ACE as a paradigm for their particular research endeavor.

ACE is ``the computational modeling of economic processes (including whole economies) as open-ended dynamic systems of interacting agents´´ (Tesfatsion 2017, p. 207). Two aspects constitute ACE. First, ACE builds on the agent-based modeling paradigm (ABM), which is not only employed in economics but also in other domains of the Social Sciences (for overviews, e.g., Squazzoni 2010, Squazzoni, Jager, and Edmonds 2014). ABM constructs ``artificial´´ agents (i.e., software agents) whose behavior follows specified rules and who interact with each other in their environment. ABM advocates a bottom-up perspective where, by means of simulation, the macro-behavior (i.e., at the `` system's level´´) emerges over time from the micro-level (i.e., agents' behaviors and their local interactions) (Bonabeau 2002, Epstein 2006b). Second, ACE comprises some deviations from more established schools of economic thought like neoclassical or new institutional economics. This second feature, in particular, results in relaxed assumptions about economic agents and, in consequence, allows for integrating findings from other scientific domains, as captured in behavioral economics, neuroeconomics, and experimental economics (Chen and Wang 2011). Therefore, compared to analytical modeling, ACE brings not only more realistic assumptions that increase external validity, but also a solution technique to deal with the difficulty of reaching tractable and closed-form solutions.[2]

---

[1] Examples are the studies of Leitner and Behrens (2013) and (Leitner, Brauneis, and Rausch 2015) on hurdle-rate mechanisms in capital budgeting, Lorscheid and Meyer (2017) on mechanisms for truthful reporting in budgeting or Wall (2017) on the learning-based emergence of incentive schemes.

The structure of the paper is as follows. We begin with an outline of research areas in management accounting which, as we believe, could particularly benefit from applying ACE. In the third section, we provide an overview of ACE. This section can be skipped by a reader familiar with the method. The fourth section proposes a framework which researchers in management accounting may employ when considering ACE as a research approach for their endeavor. The fifth section relates ACE to other research approaches frequently employed in MAR, whereby particular focus is put on a comparison of ACE to (``non-ACE´´) simulation-based approaches, closed-form analytical methods, laboratory experiments, fieldwork and to archival and survey-based research. The final section concludes with some final remarks on ACE in the field of management accounting.

For the researcher interested in adopting ACE, the accompanying internet appendix outlines the two most widely applied structured simulation approaches in ACE (part A), provides supporting information on the implementation of agent-based models (part B), a guidance for design choices which the researcher in management accounting faces when employing ACE (part C) and outlines two examples of ACE-inspired studies in MAR (part D).

---

[2] This comes at the cost of lower internal validity, as it is no longer fully clear form solving a model what exactly drives the results.

# PROMISING MANAGEMENT ACCOUNTING RESEARCH AREAS FOR THE APPLICATION OF ACE

To give the reader an idea of the broad applicability of ACE in MAR[3], we use the articles published on the occasion of the nearly coinciding 25th anniversaries of two journals in the field, namely ``*Journal of Management Accounting Research*´´ and ``*Management Accounting Research*´´. In 2015 and 2016 – on the occasion of their respective anniversaries – these journals published articles by renowned scholars in the field which reflect on the state and directions of management accounting.[4]

*1) Decision-making, errors and biases*

Given that in ACE the agents and their (inter-)actions are in the very core of modeling, a starting point for the endeavor in this section is to ask which actions are in the focus in the aforementioned `anniversary issues´, and who carries them out. According to Krishnan (2015), ``Management accounting is the study of decision making within the organization´´ (p. 188). In a similar vein, Labro (2015a) suggests to refocus MAR towards decision-making compared to other topics like performance measurement or compensation. Labro points not only to the fact that decision-making is the daily practice of managers who expect to be supported by decision-facilitating information. Moreover, in the example outlined (Labro 2015a, meanwhile in Hemmer and Labro 2019), it becomes apparent that it is not necessarily behavior like earnings management that is at hand, but – more `natural´ – managers who learn from the past over time about their subject of decision-making. In this line, Bloomfield (2015) stresses errors in managerial reporting and biases in decision-making as relevant issues for MAR. In a similar vein, Shields (2015) advocates to integrate perspectives of behavioral economics and psychology to gain a richer understanding on how management accounting affects the various components of individuals' utility. These considerations are line with Bromwich and Scapens (2016) who argue that parts of MAR rely on ``restricted assumptions´´ (p. 7) when studying topics like moral hazard or adverse selection. Against this background, we argue that ACE, with its focus on agents and broadness regarding behavioral assumptions, may contribute to study aforementioned issues. In this line, for example, ACE could be a suitable approach for the following research questions:

- How does decision-making behavior of managers which is affected by a multiplicity of biases (as elaborated in psychology and behavioral economics) impact performance obtained at an organizational level? Do certain biases mitigate or aggravate each other?

- How does heterogeneity among decision-making managers in respect to biases affect performance obtained at an organizational level?

- Do errors and noise in decision-facilitating information propagate across organizations employing division of labor? What fosters eventual propagation? Are there efficient `barriers´ against propagation?

*2) Management control systems and intra-system interactions*

---

[3] Some other areas in accounting like taxation (e.g., Davis, Hecht, and Perkins 2003, Bloomquist 2006, Bloomquist 2011) or compliance (e.g., Davis and Pesch 2013) have also used ACE, but we are particularly enthusiastic about its potential in managerial accounting research.

[4] Regarding the JMAR, issue 1 in 2015 comprises the papers of the 25th anniversary celebration with an introduction by Mittendorf (2015) and an overall reflection by Krishnan (2015); regarding ``*Management Accounting Research*´´, the 25th anniversary issue was published in volume 31 (March 2016) with an editorial article by Bromwich and Scapens (2016).

Great emphasis is placed on management control systems (MCS) being of high relevance to MAR in both `anniversary issues´ (Salterio 2015, Shields 2015, Otley 2016). According to Otley (2016), the design and use of MCS ``has generally been treated in a fragmentary manner with just one or two aspects selected by each reported study, with very little work attempting to gain a holistic view of the overall systems in use by an organization.´´ (p. 54). This broadly calls for studying different components of MCS in interrelation with each other – with the idea of some internal fit or consistency among components (e.g., Grabner and Moers 2013). However, as Luft (2016) points out based on experimental evidence, the formal and contractual components of MCS may play a double role in affecting behavior: on the one hand, by enforcing behavior which they are designed for; on the other hand, by ``influencing the behavior they cannot enforce´´ (p. 84), for example, by affecting the atmosphere in terms of (dis-)trust or informal reciprocity which may be directed towards or against organizational goals.

We argue that ACE is particularly suitable to study MCS from a comprehensive or – according to Otley (2016) – `holistic´ perspective. ACE allows to model organizations with rich institutional arrangements in which decision-making agents with multi-facetted behaviors and preferences reside. Moreover, it is an innate property of ACE that agents interact with each other where agents' actions (decisions) are the main trigger of events. Actions and interactions may be affected by MCS in its double role according to Luft (2016).

Hence, with ACE, for example, the following research questions may be studied:

- Which configurations of MCSs have a high internal consistency in terms of a high internal fit among subsystems?

- Are there subsystems / components within MCS that are more critical than others with respect to the performance regarding the overall goal of an organization?

- How do interactions among agents with heterogeneous qualifications and roles, e.g., managers and management accountants operating in an MCS affect the evolving properties of an MCS?

**OVERVIES OF AGENT-BASED COMPUTATIONAL ECONOMICS (ACE)**

The purpose of this part is to briefly characterize the key features of ACE. A reader familiar with ACE can skip this section. We structure this section into three parts: first, for a theoretical perspective, an overview of assumptions in ACE which deviate from more mainstream schools of economic thought is given; from a more `technical´ perspective, second, agent-based modeling as the modeling approach employed in ACE is outlined – followed, third, by some remarks on agent-based simulation.

**Theoretical Assumptions Incorporated in ACE**

This part intends to characterize key aspects in which ACE deviates from more traditional schools of economic thought like neoclassical economics and neo-institutional views. It is worth mentioning that there are several more traditional economic approaches employing `agent thinking´ like, for example, principal-agent theory in neo-institutional economics with its applications in understanding and designing incentive schemes. However, we seek to determine the more distinctive features of ACE as compared to more traditional schools of economic thought like the neoclassic approach and new institutionalism. To this end, we particularly rely on Tesfatsion (2006), Chang and Harrington (2006), and Chen (2016), and the key differences are summarized in Table 1.

**<<<insert Table 1 here >>>**

# TABLE 1

**Key Elements of Neoclassical and new-institutional Economics vs. Agent-based Computational Economics based on Tesfatsion (2006), Chang and Harrington (2006), and Chen (2016)**

| Components/ Aspect | Neoclassical and new-institutional economics | Agent-based computational economics |
|---|---|---|
| **Agents' search behavior** | agents are capable of overseeing the entire solution space; for given preferences and a given state of information the optimal solution is identified<br>→ optimization | agents are not capable of overseeing the entire solution space; i.e., in a given situation (current states, preferences,…) agents search stepwise for better solutions<br>→ adaptive search |
| **Agents' rationality** | rational behavior; imperfections in behavior result from incomplete information (e.g., about future states of relevant variables) | bounded rationality; cognitive limitations may cause imperfect behavior; even information that is available may not be used |
| **Agents' learning** | based on understanding and foresight | predominantly based on experience and retrospection |
| **Agents' diversity** | prevalence of the so-called ``representative agent´´, i.e., choices of an individual are regarded as coinciding with the aggregate of the agents of that type | heterogeneous agents, i.e., agents may differ with respect to various properties, which is explicitly considered |
| **Focus of analysis** | focus on equilibria with detailed analysis of properties of equilibria | focus on processes and interactions; detailed specification of structural conditions, institutional arrangements, and behavioral dispositions |
| **Time horizon** | predominantly steady state, equilibrium | after some adaptation but before system has settled |
| **Method** | closed-form models with (mathematical) proof of results | numerical experiments (simulation) |

Compared to more traditional schools of economic thinking, the presumably most distinctive feature of ACE is its view of economic agents. In particular, in ACE some of the key assumptions regarding economic agents' characteristics (which are usually made in traditional economic approaches) are relaxed, in order to obtain more realistic models. The relaxation of assumptions partially has far-reaching consequences for the models used in ACE, the way of `solving´ them and the insights provided. Given our focus on MAR, in the following, we outline those deviations of ACE from more traditional schools of economic thought which are particularly relevant from an intra-organizational perspective (for broader overviews, e.g., Tesfatsion 2006, Chen 2016). The relevant key characteristics are: (1) the understanding of agents showing some form of bounded rationality which results in (2) a focus on search and learning processes and (3) the explicit recognition of agents' heterogeneity.

*1) Bounded rationality of agents*. In neoclassical and neo-institutional economics, an agent is typically characterized by a utility function which captures the agent's preferences, and by the agent's beliefs (in terms of distribution functions, which represents rational expectations about the unknown). On this basis, roughly speaking, an agent decides in favor of that option that maximizes its expected utility function. Moreover, economic agents are assumed to understand how other agents behave. However, understanding other agents does not necessarily imply that they are completely informed about their fellow agents' behavior. Rather, there might be some form of information asymmetry, which may result in hidden action or hidden information problems (Lambert 2001).

In contrast, in ACE agents are typically characterized by some form of bounded rationality in terms of limitations of cognitive capabilities (for a conceptual overview Chen 2012). These limitations are often inspired by works of Herbert Simon (1955, 1959), and may be based on behavioral economics, neuroeconomics or experimental economics (Chen 2016).

In particular, in ACE, agents may only dispose of limited memory and limited information-processing capacity, show biases when forming expectations and employ heuristics in decision-making (Tversky and Kahneman 1974). Cognitive limitations provide the basis for fine-tuning the skills of agents comprised in an agent-based model and allow capturing less `gifted´ agents than usually suggested by the traditional schools of economic thought (Axtell 2007, Chang and Harrington 2006).

*2) Stepwise search and learning.* With agents being endowed with bounded rationality, further consequences regarding the type of `decision-making algorithm´ arise, which, in consequence, lead to another deviation from traditional economic thought about agents: As mentioned above, agents are usually modeled as utility maximizers, in principle, able to oversee the entire space of feasible solutions `at once´. Hence, agents, first, dispose of an understanding (in terms of a mental model) of the decision problems they face and, second, are endowed with foresight on what is going to happen. This enables agents to employ optimization algorithms which allow identifying the global maxima of their respective decision problems instantaneously.

In contrast, in ACE agents usually do not know the entire space of feasible solutions. It is often assumed that, at a certain time, an agent does not dispose of a `theoretical´ understanding of its problem and the possible solutions to it, but has to search stepwise for superior solutions, e.g., solutions that provide satisfyingly better performance than the status quo (with respect to the agent's objectives) (Safarzyńska and van den Bergh 2010). A rationale behind this deviation of ACE from more traditional economic thought is a satisficing approach: when information-processing capabilities are limited, agents may strive for ``good enough decisions with reasonable costs of computation´´ (Simon 1979, p. 498).

For representing such stepwise and local search behavior, predominantly greedy algorithms and, in particular, hill-climbing algorithms are employed. According to Cormen et al. (2009, p. 414), a ``greedy algorithm always makes the choice that looks best at the moment´´ in terms of ``a locally optimal choice in the hope that this choice will lead to a globally optimal solution´´. Hill-climbing algorithms employ the metaphor of seeking the highest summit in a landscape which is not entirely known to the hiker, but has to be discovered stepwise. With hill-climbing algorithms a move in the `landscape´ requires that the outcome (`altitude´) is increased by that move. For example, with a steepest ascent hill-climbing algorithm the one option of a limited set of options discovered is selected which provides the highest improvement in outcome compared to the status quo; if none of the alternatives promises an incline in outcome the status quo is kept. Therefore, hill-climbing algorithms are particularly prone to get stuck in local maxima, ridges or plateaus in a landscape (Macken, Hagan, and Perelson 1991, Cormen et al. 2009). This is, in particular, due to the fact that with these algorithms a decision-maker (`hiker´) would not accept a short-term decline in favor of a long-term increase, since no choice in favor of an option that provides an inferior outcome than the status quo would be made (Selman and Gomes 2006, Tracy et al. 2017).

Because stepwise search for superior solutions – instead of `instantaneous´ optimization – is a key feature in ACE, further questions come into play: How is search organized among a group of decision-making agents (e.g., in parallel or sequentially)? How do agents learn to search in terms of improving their search strategies? Which search strategies evolve over time (e.g., switching from search based on experience and experiential to forward looking search based on a mental model of the decision problem to be solved)? The latter aspect particularly indicates that search behavior may be subject to adaptation.

With the shift from, roughly speaking, `instantaneous´ discovery of the global optimum to stepwise search and learning the processual perspective including questions of performance enhancements in time and speed of adjustments come into play as well.[5] In comparison to neoclassical and neo-institutional economics, ACE brings along a shift from (comparing and analyzing) equilibria and long-term perspectives to emergent solutions and mid-term dynamics (Chang and Harrington 2006).

*3) Heterogeneity of agents.* The explicit consideration of boundedly rational economic agents (see above) allows for heterogeneity of economic agents, since there are numerous forms of departing from rationality (Axtell 2007, Chen 2012). Agents may, for example, be heterogeneous with respect to preferences, expectations, knowledge, cognitive abilities, basic beliefs, social network, and behavioral rules (for an overview see Chen 2016). The – potentially extreme – heterogeneity of agents in ACE is in contrast to what is called the `representative agent´ (Kirman 1992) which is common in more traditional economic models. Even if more traditional economics considers different types of agents (e.g., consumers, firms) the numerous agents of one type are represented by one `representative´ individual whose choices are regarded to coincide with the aggregate of the agents of one type (a survey on aggregation in economics is given by Blundell and Stoker 2005).

The set of assumptions incorporated in ACE comes at the cost of mathematical tractability compared to more traditional economic approaches: rather than closed-form models which allow for proofs of optimal solutions or equilibria (e.g., in principal-agent models for optimal incentive schemes), ACE carries out numerical experiments using computational models and simulations. In the following, agent-based modeling as the particular computational modeling approach employed in ACE is outlined.

---

[5] We return to these particularly time-related perspectives of ACE within the following sections of this article.

**Agent-Based Modeling (ABM)**

Agent-based models are composed of three key ingredients: (1) agents, (2) the environment in which the agents operate, and (3) interactions among agents, and among agents and environment (e.g., Tesfatsion 2006, Macal and North 2010). The researcher models a system consisting of these components, and via computer simulation, `histories´ of the agents' interactions (micro-level) and the emergent system's behavior (macro-level) are generated (Tesfatsion 2017). For this reason, ABM is assumed to lead to a `generative social science´ (Epstein 2006a). In the following, the three key components of agent-based models[6] are described in more detail.

(1) *Agents* reflect the basic entities which drive – by their actions and decisions – the dynamics that enfold the `histories´ generated by the simulations. The real-world entities which are described in terms of agents depend on the particular focus of the research endeavor. For example, agents may represent

- single individuals, like managers, workers, auditors or board members

- social groups, i.e., groups of persons with a similar professional background like accountants or with a similar history in the firm (e.g., same cohort or hiring date)

- institutions, like markets, firms, departments within an organization

---

[6] It is worth mentioning, that there are two major streams of `agent-based computing´ which should be clearly distinguished – though there are commonalities and cross-fertilizing interrelations between the two (Niazi and Hussain 2011): on the one hand, there is agent-based modeling and agent-based computing employed for the *understanding* of emergent phenomena in complex systems of interacting agents. This stream of research employs agent-based modeling as a research method, often termed as `social simulation´ (e.g., Squazzoni, Jager, and Edmonds 2014) and also reflects the understanding of this paper. On the other hand, agent-based modeling is applied in the domain of multi-agent systems (e.g., team of robots, fleets of drones) in order to support the *design* of systems with the intended specifications (Wooldridge 2009).

- agents of other `species´ such as robots or autonomous decision-making devices based on methods of artificial intelligence

Agents are endowed with some basic characteristics such as the objectives which they pursue, their possible actions or decisions to achieve these objectives, and their behavioral rules (e.g. Bonabeau 2002, Safarzyńska and van den Bergh 2010, Tesfatsion 2006, Wooldridge and Jennings 1995). Regarding the latter, it is an essential characteristic of agents that they have some autonomy, meaning that their behavior is not directly and completely determined by a central authority. In other words, without endowing an agent in the model even with a slight behavioral discretion, this agent cannot contribute much to the understanding of the emergent behavior at the macro-level. However, this does not mean that no hierarchical levels are included in agent-based models. Interactions including feedback loops between subordinate decision-makers and superiors could, for example, be included in ABMs in order to represent hierarchical organizations (Epstein 2006a, Chang and Harrington 2006).

Moreover, in ABM agents necessarily dispose of capabilities (i) of receiving information from both their environment and fellow agents, and (ii) to respond to the information received. In a model on the diffusion of innovation this could, for example, include that agents (i) observe whether their neighbors have already adopted a new product and (ii) decide in favor of that product, too, if a certain threshold of adopting neighbors is reached. Hence, in an agent-based model, an agent is represented by a set of data on the agent's particular knowledge about the environment and other agents, and a set of methods which capture the agent's behavior; both agents' knowledge and behavior may be subject to change over time (Tesfatsion 2006, 2017).

(2) The *environment* in which the agents operate, constitutes the second building block of agent-based models. The term `environment´ is broad and – depending on the subject of the

model – could address a physical space (e.g., geographical locations and technical infrastructures), the ecological environment (e.g., dispersion of plants and atmospheric pollution) or, presumably more relevant for management accounting, a conceptual space. In MAR, such a conceptual space could, for instance, be given by the tasks or decision problems which the agents face and the constraints to be respected when fulfilling tasks (Chang and Harrington 2006).

Irrespective of the broad understanding of `environments´, two aspects for the modeling should be stressed. First, in ABM the environment is usually *explicitly* represented, (e.g., the particular spatial environment including topology, streets) (Epstein 2006a). In organizational modeling, where the environment may represent the decision problems to be solved, this could mean that the entire solution space including the performance levels provided by the solutions is represented. This justifies the term `fitness landscape´, which is often used in ABM and is inspired by evolutionary biology (Kauffman and Levin 1987, Levinthal 1997). However, the agents operating in a `landscape´ usually do not know it entirely and, instead, have to explore it (as mentioned in the previous section). Second, the explicit modeling of the environment allows determining whether two agents or two solutions are `*neighboring*´ each other (either literally in spatial models or in a figurative sense, e.g., regarding cultural distance). Based on the idea of neighborhood, different search strategies could be studied, for instance, whether agents search `locally´ (i.e., only slightly modify their current solution which refers to exploitation) or execute `long jumps´ (i.e., make major changes which corresponds to exploration) to find superior solutions to their concerns.

The environment may show different characteristics. For example, it could represent simple or highly complex tasks that are or are not decomposable (Simon 1962, Chang and Harrington 2006). The environment may be stable or of dynamic nature which could be exogenously given (e.g., shocks) or brought about endogenously when the environment is shaped by (or reacts to) agents' actions (e.g., Siggelkow and Rivkin 2005).

(3) *Interactions* among agents, and between agents and their environment represent the third core component of agent-based models. The *communication* among agents may be represented by direct and/or indirect interactions (Safarzyńska and van den Bergh 2010, Weiss 1999, Tesfatsion 2001). *Direct* interaction means that agents, in whichever way, directly communicate with each other. In models on intra-organizational coordination, agents capturing subordinate managers may, for example, inform other agents in a certain sequence at the same hierarchical level about their intended actions; in consequence, these agents may revise their intentions when they receive information about intended actions (e.g., Martial 1992). Regarding feasible options, subordinate agents may inform an agent representing a higher-level manager who, then, makes a final choice (e.g., Siggelkow and Rivkin 2005). In the aforementioned lateral and vertical communication processes, communication may work perfectly or be impaired by errors (Wall 2016, 2019). *Indirect* interactions among agents are usually implemented to work through the environment by, for example, agents observing each other, noting how the environment is affected by the actions of other agents and reacting to the changes in the environment. Observing other agents may also allow agents learn from each other and, in this vein, to capture imitation (e.g., Rivkin 2000). Interactions may also be classified according to whether or not they are of cooperative or non-cooperative nature (e.g., sharing/not sharing knowledge and other resources) (Axelrod and Hamilton 1981). Moreover, ABM allows capturing more complex social interactions among agents based on friendship, sense of belonging to certain social groups or collective emotions (Schweitzer and Garcia 2010).

Hence, the modeler has to specify an underlying `topology of connectedness´ which defines with whom and how agents interact (Macal and North 2010). This also includes the definition of a time structure of interactions, e.g., (artificial) synchronicity versus asynchronicity (Tesfatsion 2017). In case the `topology´ of interactions is subject to (endogenous) change, rules for the adaptation of interactions have to be specified, too.

It is noteworthy, that in ABM the behavior at the macro-level emerges from the agents' actions with their local interactions at the micro-level. Even simple behavioral rules and local interactions at the micro-level may induce remarkably complex behavior at the macro-level (e.g., Ma and Nakamori 2005) which cannot be derived in terms of a `functional relationship´ from the individual behaviors of those agents (Epstein and Axtell 1996, Epstein 2006a, Tesfatsion 2006, 2017).[7]

**Agent-Based Simulation**

Agent-based models are typically implemented in software programs and `solved´ via computer simulation, i.e. not proofs, but extensive numerical derivations produce the results (Chang and Harrington 2006).

---

[7] However, ABM is not the only modeling approach with a particular focus on understanding social systems and the dynamics of complex systems. In particular, system dynamics has to be mentioned. In system dynamics, modeling starts with assumptions about functional relations between variables capturing characteristics of the system which is the center of interest. The functional relations may be set up based on empirical evidence or theoretical assumptions. From the functional relations (often a set of differential equations), predictions regarding equilibria or the diffusion of innovation or change can be derived. A key difference to ABM is that system dynamics starts directly at the system's level, and system properties are derived from a set of equations by recording the changes of variables over time (Gilbert and Troitzsch 2005). In contrast, in ABM, at the starting point of modeling there are the agents, their behavior and interactions and the researcher observes which properties of the system emerge bottom-up from these ingredients. The two modeling approaches differ by how individuals (e.g., decision-makers) are captured: in system dynamics individuals are typically not modeled individually but captured as representative agents, i.e., each agent is similar and represents the `average´ of the population of interest. In contrast, in ABM individuals are modeled distinctly. According to Macy and Flache (2009), ABM ``replaces a single unified model of the population with a population of models, each of which is a separately functioning autonomous entity´´ (p. 255). For thorough comparison of the two modeling approaches see Schieritz and Milling (2003).

According to Law (2007), simulation models can be classified along three dimensions. First, the categorization in *static vs. dynamic models* captures whether a model represents a system at a certain point in time or whether the system is observed over time. Second, the differentiation in *continuous vs. discrete* models captures – roughly speaking – whether in the representation of the system modeled the state variables change continuously with respect to time or instantaneously at certain points in time. Third, *stochastic* models differ from *deterministic* models according to whether, or not, they contain any input components which are of probabilistic (i.e., random) nature.

According to this categorization, agent-based simulations are *dynamic* and *discrete*, and usually comprise *probabilistic* components. In particular, agent-based simulations are *discrete event simulations* with the specific characteristic that agents' actions are the main trigger of system changes (Dooley 2002), while the `simulation clock´ is advanced by a fixed-increment time mechanism (Law 2007). Often, agent-based simulations are *dynamic* according to Law's (2007) understanding in that the system can evolve over time, which allows studying processes of adaptation, diffusion, imitation or learning. Agent-based models most commonly include *stochastic* components in order to capture external shocks or limited cognitive capabilities of agents (for more information see the subsequent section).

Davis et al. (2007) distinguish between structured and unstructured simulation approaches, whereby the former have some built-in properties, while the non-structured approaches are customizable, offering more flexibility to the modeler. Agent-based models in managerial science which pursue a structured approach, mostly employ cellular automata and NK fitness landscapes. Key features of the two approaches are outlined in part A of the internet appendix, which also includes hands-on guidance on implementation of ABM.

# A FRAMEWORK FOR EMPLOYING ACE IN MANAGEMENT ACCOUNTING RESEARCH

The purpose of this section is to systematically investigate what it is that the perspective of ACE may contribute to management accounting research. We specifically focus on ACE with its built-in theoretical assumptions and modeling approach. In particular, building on the paradigmatic elements of ACE as introduced previously (i.e., bounded rationality, stepwise search and learning, heterogeneity of economic agents) and the characteristics of the agent-based modeling paradigm as outlined above, we propose a framework of five key features of ACE relevant for MAR and discuss their particular opportunities and difficulties. This framework - summarized in Table 2 - may provide an orientation for researchers in management accounting to decide whether ACE could be an appropriate approach for their research endeavor. The framework is complemented by Appendix C which provides some guidance for the design choices to be made when employing ACE for MAR.

**Realistic Behavioral Assumptions about Agents' Rationality:**

The assumptions about the agents' bounded rationality is probably the most compelling aspect of ACE and, as such, can refer to sound empirical evidence from various streams of research.

With respect to MAR, we think that this could foster two streams of research in particular – both referring to decision-making in the presence of biases and errors as mentioned in the second section. The first stream may be termed `agentization´ of economic models according to Guerrero and Axtell (2011) and Leitner and Behrens (2015). In MAR, this procedure could be particularly relevant for the design of incentive schemes and (tests of the incentive-compatibility of) performance metrics. The idea behind `agentization´ is to systematically relax assumptions about economic actors included in economic models, and to study the conditions under which the optimal solutions (e.g., optimal contracts) proposed by classical economic models hold in contexts with less gifted agents (e.g., Leitner and Wall 2019). In this sense, `agentization´ may be regarded as an approach for robustness analysis as it allows to study management accounting in the context of more realistic decision-makers than reflected in more traditional schools of economic thought. With respect to MAR, this may above all refer to research on incentive schemes as this stream of research is predominantly based on the principal-agent theory and, therefore, includes assumptions about `gifted´ economic agents. These restrictive assumptions may give reason to the limited practical relevance of this research, as diagnosed by Bromwich and Scapens (2016). Example 1 introduced in part D of the internet appendix presents the outline of an `agentized´ version of the standard hidden-action model (Holmström 1979).

A second aspect of management accounting, which we think could particularly benefit from taking bounded rationality into account, is related to accounting information in managerial decision-making and its effects at an organizational level. This refers to behavioral streams of accounting research (Hopper and Bui 2016). Given the broad evidence from behavioral experiments on the diverse biases and heuristics in decision-making, it is of interest to study how such effects shape the performance of management accounting practices in organizations. If, for example, various interacting agents suffer from (potentially different) biases it is interesting how accounting-based information may shape the overall performance obtained by an organization.

The two aforementioned `applications´ of ACE in MAR build on the higher flexibility with respect to the modeling of economic actors that ACE grants to the researcher, as compared to more traditional schools of economic thought. At the same time, due to its computational nature, ACE requires to precisely specify – in a mathematical sense – the boundedness of agents' rationality.

However, the flexibility gained by abandoning the idea of (perfectly) rational economic agents comes at some difficulties for MAR. As such the peril of a lack of rigor is to be mentioned, or as Axtell (2007, p. 107) puts it: ``…there is one way to be rational but many ways to depart from rationality´´. Since there are various forms of bounded rationality (for an overview of modeling history regarding bounded rationality we refer to Chen 2012), arbitrariness could be an issue. In a similar vein, modeling agents with a realistic level of ``intelligence´´ seems to be a challenge. In particular, modeling agents with limited cognitive capabilities – like the prominent `zero-intelligence agent´ with its surprising predictive power for financial markets (Farmer, Patelli, and Zovko 2005) – is not difficult. However, ``to extend their intelligence to the point where they could make decisions of the same sophistication as is commonplace among people'' (Gilbert 2008, p. 16) turned out to be a challenge in agent-based modeling.

**Rich Institutional Arrangements and Rich Interactions**

Management accounting is embedded in a broader context of institutional arrangements such as the hierarchical structure of organizations or the management control system (Chenhall 2012). Management Control Systems (MCS), intended to align the behavior of organizational members with an organization's objectives, incorporate various controls like rules and behavioral norms, performance measures, incentives and feedback mechanisms to name but a few (for an overview of conceptualizations of management controls see Langfield-Smith 2006). The agent-based modeling paradigm could capture rich configurations of institutional arrangements as comprised in MCS. In this context, it is also worth mentioning that ABM allows for representing organizations with various hierarchical levels, numerous decision-making agents, and rich interactions among them as, for example, shaped by the MCS (Chang and Harrington 2006).

Accordingly, ACE may contribute to the recent debate on the internal fit of management controls in organizations: management controls may be more effective when they form an aligned system rather than a collection of techniques (Grabner and Moers 2013, Malmi 2013, Bedford, Malmi, and Sandelin 2016). Agent-based models could be employed to systematically explore configurations of management controls, and, thus, build a bridge between research based on closed-form models with their reduced institutional settings and empirical research in that domain (e.g., Bedford and Malmi 2015).

However, this high flexibility incorporated in ACE with its agent-based modeling paradigm also bears some difficulties which, to some extent, correspond to those mentioned before. As such, behavioral rules in ABM are often introduced ad hoc and lack sufficient justification, which has led to calls for more common frameworks in ABM, particularly in economics (Richiardi et al. 2006, Safarzyńska and van den Bergh 2010).

**Rich Contingencies**

There is long tradition of contingency-based research in management accounting. One aim of this stream of research is to study which configurations of MCS fit their context (based on, for example, culture, environmental turbulence, age, size), where fit is meant as the performance effect resulting from the combination of MCS design and context (Chenhall 2003, Grabner and Moers 2013).

The agent-based modeling paradigm incorporated in ACE allows representing rich sets of contingency factors, which could provide a promising correspondence to contingency views in MAR. This may be most obvious with respect to the `environment´ which is one of the key ingredients of agent-based models and is usually explicitly modeled, as mentioned above. For example, the environment may reflect the complexity of the task an organization faces due to the level of interdependencies among subtasks. Other forms of contingencies could be conveniently captured in an agent-based model (e.g., skills of agents, cultural distances among agents, level of environmental turbulence) and, in the end, what is regarded as contingency factor depends on the particular experimental design.

Consequently, the perspective of ACE may contribute to MAR in several ways: First, it may build a link between theory-creating and theory-testing approaches (Davis, Eisenhardt, and Bingham 2007) by generating hypotheses based on simulation results which could ideally be empirically tested. Second, the effect of contingencies for which empirical data are hard to obtain could be studied via (agent-based) simulation. This could apply, for instance, to the effects of errors in costing systems, which are hard to figure out empirically (Labro and Vanhoucke 2007, Leitner 2014). Third, due to its ability to systematically control for a variety of contingencies, ACE might be an indicator of interrelations among contingent factors, in terms of combinations that affect the performance of MCS.

However, given the usually vast number of variables in simulation models, it may be tempting in this case to examine too many settings of variation – at the risk of getting lost in the huge combinatorial space, resulting in difficulties to come up with concise and transparent results (Labro 2015b).

**Processual Perspective: Learning and Emergence**

As outlined above, the particular assumptions about economic agents included in ACE result in focusing on processes of adaptation: agents' stepwise search (in order to adapt towards performance at higher levels) takes place and new behavioral rules and interactions may emerge, which requires to endow agents with some of the various kinds of learning.

We argue that the processual perspective incorporated in ACE could contribute to MAR in various ways. First, assuming that economic agents are not (perfectly) rational raises the question of how fast they identify superior solutions and how a given set of management controls (i.e., incentives provided) affects the *speed* of performance enhancements (Baines and Langfield-Smith 2003). Second, under the assumptions of ACE, management controls themselves may be subject to learning. For example, in contrast to the principal-agent-theory, the principal has to *learn* about the set of feasible incentive schemes and their contributions to performance, which is illustrated in the first example of the subsequent section. Third, ACE allows studying the *coevolution* of management controls and changes in contingencies: this refers, for instance, to the stream of MAR which studies the change of MCS in the course of firm growth, as initiated by the seminal paper of Davila (2005) and illustrated by the second example in part D of the internet appendix. With this ACE meets calls for fostering a dynamic perspective in MAR, particularly with respect to a contingency approach since it is at the very core of this approach to focus on processes of how organizations adapt over time in response to changing contingencies (e.g., Hall 2016, Otley 2016).

However, endowing a system captured in an agent-based model with `dynamic capabilities´ so that adaptation and emergence can happen requires the researcher to make difficult design choices. For example, to let the MCS evolve with firm growth some form of learning may be employed and, consequently, the researcher has to make a choice among various forms of learning (e.g., Brenner 2006, Chen 2012), which could affect the dynamics in the simulation experiment considerably.

**Bridging the Micro-Macro Divide**

In order to gain a deeper understanding of the functioning of organizations, it is necessary to consider the links between both the micro- and the macro-layer of systems under investigation (e.g., organizations) (Molloy, Ployhart, and Wright 2011). The micro-level can, for instance, refer to the individual perspective or groups (e.g., in terms of goal setting) and the macro-level refers to the overall organizational level (e.g., in terms of firm performance or corporate strategy) (Klein, Tosi, and Cannella Jr 1999, Huselid and Becker 2011).

Establishing this link has long been challenging for researchers (Gavetti 2005, Teece 2007).[8] The focus on either the micro- or the macro-level has led to a relatively large set of level-specific theories and a level-specific terminology. Finding a common language and a common theoretical basis among researchers who focus either on the micro- or the macro-level might be challenging. Also, researchers at both the micro- and the macro-level are familiar with their `own´ view on the problem under investigation: Psychologists might have a special interest in the effects at the individual level, while some economists are particularly interested on the macro-level. Connecting the two levels might, thus, be challenging as it is in the nature of ACE that macro-level effects *emerge* from micro-level interactions and behavior. Establishing this link potentially results in complicated interactions between the two levels and, therefore, contributes to the observed phenomenon that simulation models are often perceived as `black boxes´ (Lorscheid, Heine, and Meyer 2012, Labro 2015b). An integrated view (or a `grand organization and management theory´, which Molloy, Ployhart, and Wright (2011) refer to) requires to link the two layers. There are responses to the micro-macro divide. First, there is the approach to explain phenomena at the macro-level by what happens at the micro-level (Gavetti 2005, Teece 2007, Powell and Colyvas 2008). This first response lies at the heart of ACE. For research in management accounting, a proper use-case can be a system in which individual actions and choices are shaped through incentive schemes. ACE allows observing individual responses to different types of management controls, interactions among management controls as well as among individual actions and decisions (i.e., the micro-level), and their effect on organizational performance (i.e., the macro-level). From a management control perspective, this allows for linking research on individual incentive schemes with an organization's performance management system. Second, there is a body of research which aims to set up multi-level theories. The latter approach focuses on relationships between levels and similarities in structures and processes across levels (Kuhn 2012). Although responses exist, the micro-macro divide is still a predominant challenge for management (accounting) research (Delery and Roumpi 2017).

---

[8] Note that the divide between research at the micro- and the macro-layer, respectively, is not a phenomenon specific for management accounting research only, but for the social sciences in general (Kuhn 2012). It is in the nature of some academic fields to have a strong focus on either the micro- or the macro-level, which is why the divide is sometimes also referred to as *disciplinary divide* (Bargiela-Chiappini and Nickerson 2002). A large proportion of research carried out in the field of Psychology, for example, focuses on individuals (i.e., the micro-level). Other economic fields have a strong focus on the macro-level: In Economics, for instance, research often focuses on the level of institutions or the level of economic systems.

Table 2 summarizes these opportunities and difficulties of key features of ACE in MAR with the difficulties listed in the far-right column predominantly resulting from the high flexibility that ACE and ABM grant to the researcher.

Recently, considerable progress was made in order to provide researchers employing simulation with guiding principles and frameworks to increase transparency and validity of results (Grimm et al. 2006, Janssen et al. 2008, Lorscheid, Heine, and Meyer 2012). Part C of the internet appendix introduces core design choices to be made by a researcher for employing agent-based modeling in MAR according to the ODD+D protocol (Müller et al. 2013).

**(<<<< insert Table 2 >>>>)**

# TABLE 2

# Opportunities and Difficulties of ACE in Management Accounting Research

| Key Features of ACE | Opportunities | Difficulties |
|---|---|---|
| 1) Realistic behavioral assumptions about agents' rationality | - more realistic representation of economic agents in MAR though precisely specified for computational implementation<br>- potential positive effects on practical relevance of MAR<br>- particularly relevant for<br>  - `agentization´ of traditional economic models in MAR, especially in the context of incentive schemes<br>  - study of effects of accounting information in decision-making in organizations | - potential arbitrariness of modeling agents with bounded rationality<br>- realistic levels of agents' boundedness / sophistication hard to model |
| 2) Rich institutional arrangements and rich interactions | - representation of multi-level organizations resided by numerous agents<br>- representation of a multitude of controls incorporated in MCS, for example, via behavioral rules<br>- particularly relevant for analyzing the internal fit among the controls incorporated in MCS | - potential arbitrariness related to interactions and behavioral rules captured by the model |
| 3) Rich contingencies | - correspondence to contingency-based MAR<br>- representation of a multiplicity of contingent factors of MCS for generating empirically testable hypotheses<br>- particularly relevant for<br>  - analyses of contingencies that are hard to track empirically<br>  - systematic exploration of interactions among contingencies | - huge combinatorial space of parameter settings regarding contingent factors which may hamper obtaining concise and transparent results |
| 4) Processual perspective: learning and emergence | - representation of longitudinal and processual phenomena in MAR<br>- focus on mid-term dynamics and turbulence (rather than steady state and equilibria)<br>- particularly relevant for<br>  - study of effects of MCS on speed of performance enhancements<br>  - analysis of intra-organizational learning of management controls and the design of MCS<br>  - analysis of emerging properties of MCS in coevolution with contingent factors | - critical choices to be made by the modeler on how to `induce´ dynamics regarding, in particular, the type of search employed by agents and the mode of learning |

with ACE: agent-based computational economics; MAR: management accounting research; MCS: management control system

# ACE IN COMPARISON TO AND IN COMBINATION WITH OTHER RESEARCH METHODS IN MANAGEMENT ACCOUNTING

This section serves to highlight the pros and cons of ACE compared to other research methods in management accounting and to present some considerations on how ACE could be fruitfully combined with other methods in MAR. The aim is to provide the researcher with some decision-making aid as to whether or not ACE could be a promising approach for a particular research question.

To this end, we structure our line of arguments according to the following considerations. ACE in its `technical core´ relies on *simulation* as a research method. Hence, the strengths and weaknesses inherent in simulation-based research reasonably also apply to ACE. Moreover, ACE and agent-based modeling in particular, primarily aim at *theory-building*. Both aspects are briefly discussed in the first sub-section.

Therefore, when, in the second sub-section, comparing the pros and cons of ACE against other research methods in management accounting we take the perspective of theory-building. We compare pros and cons of ACE to (non-ACE) simulation-based, analytical, experimental, fieldwork, archival and survey research with respect to construct, internal and external validity.

Third, complementing theory-building by ACE with *theory-testing* methods is a `self-evident´ combination. However, we believe that there are further promising applications for combining ACE with other methods in MAR. In the third sub-section, we briefly outline some examples with particular focus on empirical methods. Moreover, we illustrate how incomplete contracts could be a promising topic for combining ACE with other methods.

## On the Methodological Dimension of ACE

### *Some Remarks on Simulation as a Research Method*

Agent-based computational economics typically employs simulations to `solve´ the models and, in this sense, the particular opportunities and limitations of simulation-based research in general apply to ACE, too. Comprehensive discussions on simulation as a research method and its particular strengths and weaknesses in the social sciences are, for example, given by Axelrod (1997) or Lorscheid, Heine, and Meyer (2012), in the context of managerial science by Davis, Eisenhardt, and Bingham (2007) and Harrison et al. (2007). Labro (2015b) discusses the difficulties of applying simulation methods in accounting research by reflecting on her own simulation-based research on cost accounting (Labro and Vanhoucke 2007, 2008).

Out of these surveys and discussions the following strengths and opportunities of simulation-based research appear to be particularly worth mentioning in our context: First, simulation models allow capturing ``complex multilevel, and mathematically intractable phenomena´´ (Harrison et al. 2007, p. 1240). Second, simulation requires putting the subject to be studied into a precise formal representation to be an executable piece of software (Leombruni and Richiardi 2005). This is why *internal validity* and *construct validity* are regarded as particular strengths of simulations (Davis, Eisenhardt, and Bingham 2007). Internal validity, i.e., the level to which causalities among independent and dependent variables can be inferred is enforced by the algorithmic representation in the software (Carroll and Harrison 1998). In a similar vein, the computational representation requires the researcher to precisely specify constructs and their measures which refers to construct validity, i.e., the degree to which constructs employed in the model capture what they are intended to and are measured accurately (Embretson 1983). Third, according to Davis, Eisenhardt, and Bingham (2007) simulation is especially useful for *bridging between theory-creating and theory-testing research,* for example, between case studies or closed-form modeling on the one hand, and multivariate statistical analysis on the other. Fourth, simulation was found to be particularly useful for studying processual and longitudinal phenomena which could pose considerable problems with respect to availability of data in empirical research. Fifth, another advantage of simulations is that they can reveal `*out-of-normal´ situations*: For example, simulation allows exploring non-linear phenomena like tipping points or disaster scenarios. Sixth, in a similar vein, the boundary conditions of empirical findings could be hard to determine empirically due to low availability of data for more extreme conditions. In contrast, experimenting with extraordinary parameter settings could usually be done easily once the model is implemented. This however bears the peril that extraordinary parameter settings may be overestimated in the interpretation of results (Labro 2015b).

In this sense, simulation-based research faces several further difficulties and drawbacks in domains of managerial science. As such, aspects of external validity (in terms of generalizability to wider contexts or settings) of simulation models have to be mentioned. It may be regarded as the `other side of the coin´ of representing the subject under investigation in a precise and computationally executable form that simplification and abstraction of the real-world phenomena could lead to a lack of external validity (Davis, Eisenhardt, and Bingham 2007, Harrison et al. 2007). We address this aspect – particularly in respect to closed-form models - in more detail in a respective sub-section below. Further drawbacks of simulation-based research result from the variables employed. First, simulation models often use a variety of variables (parameters) that could be manipulated, even with a high or even infinite number of possible graduations. In consequence, the combinatorial space of scenarios potentially captured in the simulation experiments may `explode´, which, in turn, affects deriving clear and concise findings from the experiments (Lorscheid, Heine, and Meyer 2012, Labro 2015b). Second, as Labro (2015b) points out, a potential peril for external validity of simulation-based research is that the parameters employed in the simulations have to be calibrated and it might be a problem find empirically sound parameter settings. This is a particularly relevant issue since the simulation results may be subtly shaped by the parameter values in combination with each other. This leads to a further potential drawback of simulation-based research, namely, its potential lack of transparency, the so-called `black box´ property (e.g., Lorscheid, Heine, and Meyer 2012). Simulation models and results often suffer from being incomprehensible to other researchers, which may be caused by the algorithmic structures of the programs, the variety of variables and parameters employed or the opaque potential interactions among parameter settings. To overcome these difficulties – and eventually abandon the relative unfamiliarity of simulation in accounting research – more extensive and elaborate descriptions of simulation models are required (Labro 2015b). While this may conflict with page limits set by publishers (Lorscheid, Heine, and Meyer 2012) the use of online appendices may be an interesting way to provide extensive documentation of simulation models and results as, for example, in Anand, Balakrishnan, and Labro (2019).

*Focus of ACE on Theory-Building and Some Inherent Obstacles*

It has been argued that in organizational and managerial science developing a simulation model, is primarily an exercise in theory-building. It requires the researcher to identify key actors, roles and processes including interactions and to specify these components in a set of computational rules, which was argued to be an inherently theoretical endeavor (Harrison et al. 2007, Davis, Eisenhardt, and Bingham 2007). However, it is worth mentioning that simulation-based research usually does not start with a hypothesis; rather the computational results of the simulation model may be regarded as the hypotheses in terms of explanations for observed phenomena (Davis, Eisenhardt, and Bingham 2007) and/or predictions of a theory (Harrison et al. 2007).

In the social sciences, ABM widely follows this line[9]. However, due to its modeling paradigm ABM provides explanations in a `bottom-up´ way or, as Epstein – one of the pioneers of ABM – describes in an apt explanation: "If you didn't grow it, you didn't explain its emergence" (Epstein 1999, p. 43). Hence, for explaining some macroscopic patterns (e.g., effects of a new management accounting technique on a firm's overall performance), according to Epstein the researcher proceeds as follows: "Situate an initial population of autonomous heterogeneous agents in a relevant spatial environment; allow them to interact according to simple local rules, and thereby generate—or "grow"—the macroscopic regularity from the bottom up." (Epstein 1999, p. 43).

However, as Macy and Flache (2009) following Epstein point out, three related reasons should prevent the researcher from overestimating the explanative power of a certain agent-based model. First, even when a certain model generates (``grows´´) the macroscopic pattern of interest, there may be alternative models which yield the same pattern. Hence, allowing the macroscopic phenomenon to emerge via simulation this does, no more and no less, show that an explanation is possible – but not how much confidence is to be put in a particular explanation compared to its potential alternatives. Second, even if only one model can be found that allows the emergence of the macroscopic phenomenon which the researcher seeks to explain, assumptions built-in in that particular model may not be plausible (we come back to that aspect in the context of combinations of ABM with other research methods). Third, even if none of the aforementioned obstacles applies, i.e., only one model which is built on plausible assumptions produces the macroscopic pattern, this model may be too complicated. In particular, its internal dynamics may not reveal how the emerging property has exactly emerged. In other words: the causalities which produce the macro phenomenon from the micro level may be opaque. The latter aspect relates to the well-known weakness of employing a multitude of parameters in simulation-based research in general, as mentioned in the previous sub-section. At the same time, this refers to an important obstacle compared to analytical research which, among other methods, we address next.

---

[9] In some domains of computer science and engineering, agent-based technologies are primarily applied with a *design* focus to multi-agent systems, like in swarms of robots (for a deeper analysis on commonalities and differences between agent-based approaches in social sciences and computer sciences/engineering we refer to Niazi and Hussain (2011)).

**Strengths and Weaknesses of ACE in Comparison with Other Methods for Theory-Building**

In this section we compare ACE to other research methods employed in management accounting: non-ACE simulation studies, analytical modeling, laboratory experiments, fieldwork, archival and survey-based research. Non-ACE simulation studies, analytical research and laboratory experiments have more in common with ACE than the other methods, the former due to the common ``technical core´´ of simulation, analytical research since it also employs formal modeling and the latter due to its experimental nature, which is why we put particular emphasis on these methods.

In the following we seek to highlight particular strengths and weaknesses, opportunities and threats of these research methods as elaborated in the related literature compared to those of ACE. We focus particularly on construct validity, internal validity and external validity as the core criteria of research methods in MAR (e.g., Abernethy et al. 1999, Bloomfield, Nelson, and Soltes 2016, Smith 2019). Broadly speaking, construct validity refers to the degree to which constructs accurately capture what they are meant to and to whether they can be measured precisely (Abernethy et al. 1999, for a discussion Embretson 1983). Internal validity captures the degree to which causalities among independent and dependent variables can be inferred, while external validity describes to which extent the results can be generalized from the particular study to wider contexts, settings or times.

The key aspects of the subsequent comparisons of research methods to ACE are summarized in Table 3.

**<<<<< insert Table 3 >>>>>**

# TABLE 3

# Major Pros and Cons of ACE Compared to Research Methods in MAR

| a. (Non-ACE) Simulation-based research | rel. pros & cons | Agent-based computational economics | Criterion |
|---|---|---|---|
| computational representation | ≈ | computational representation | C |
| narrow contextualization w.r.t. agents' behavior: few variables and algorithms | + − | rich contextualization w.r.t. agents' behavior: manifold of parameters, sub-models, interactions | I |
| typically narrow and/or standardized behavioral assumptions | − + | rich behavioral assumptions (e.g., bounded rationality including cognitive illusions, heuristics) | E |

| b. Analytical research | rel. pros & cons | Agent-based computational economics | Criterion |
|---|---|---|---|
| inference via mathematical analysis/proof | + − | inferences from numerical examples | I |
| narrow contextualization: few variables and functional relations for reasons of tractability | − + | rich contextualization: manifold of parameters and variables, multitude of interactions | E |
| tight behavioral assumptions (e.g., utility-maximizers) for reasons of tractability | − + | rich behavioral assumptions (e.g., bounded rationality including cognitive illusions, heuristics) | E |

| c. Laboratory experimental research | rel. pros & cons | Agent-based computational economics | Criterion |
|---|---|---|---|
| inferences from manipulation of few variables (treatments) | (+/−) | inferences from variation of a multitude of parameter settings | I |
| human participants | + − | computational models of humans | E |
| narrow and often abstract contextualization | − + | rich contextualization | E |

| d. Fieldwork research | rel. pros & cons | Agent-based computational economics | Criterion |
|---|---|---|---|
| in-depth understanding | (+/−) | inferences from variation of a multitude of parameter settings | I |
| real and potentially full context | + − | artificial and rich (though) parameterized contextualization | E |
| small samples | − + | large scale samples | E |

| e. Archival research* | rel. pros & cons | Agent-based computational economics | Criterion |
|---|---|---|---|
| limited control over data and unspecific contructs | − + | full control over constructs and data output | C |
| longitudinal data possibly allowing for natural treatments | (+/−) | variation of a multitude of parameter settings; processal and longitudinal perspective | I |
| real-world and large-scale data | + − | artificial large-scale data | E |

| f. Survey research* | rel. pros & cons | Agent-based computational economics | Criterion |
|---|---|---|---|
| particular challenges in longitudinal studies; predominantly cross-sectional | − + | variation of a multitude of parameter settings; processal and longitudinal perspective | I |
| rich contextualization | ≈ | rich contextualization | E |
| real-world and large-scale data | + − | artificial large-scale data | E |

\* predominantly employed for theory-testing, not for theory-building as ACE

C: construct validity;   I: internal validity;   E: external validity

(+/−): relative pros and cons depend on the research question;   ≈ : relative pros and cons equivalent for both methods

### *(Non-ACE) Simulation-Based Research compared to ACE*

Though simulation (often called numerical experiments in MAR) was relatively rarely employed in MAR compared to other methods (e.g., Balakrishnan and Sivaramakrishnan 2002, Hesford et al. 2007) there is a tradition of (non-ACE) simulation in MAR. For example, simulation was employed in research on cost accounting with respect to errors in costing systems (Labro and Vanhoucke 2007, Dhavale 2007, Leitner 2013), precision of cost-based information (Banker and Hansen 2002, Hoozée and Hansen 2018) or time-delayed information about actual cost (Anand, Balakrishnan, and Labro 2017). Simulation was applied at the interface between operations management and management accounting (O'Brien and Sivaramakrishnan 1996, Balakrishnan and Sivaramakrishnan 2002, Dhavale 2005, Leitch, Philipoom, and Fry 2005, Bai, Kajiwara, and Liu 2016). Schroeder (1992) applies simulation-based methods to a multi-period setting on the principal's choice of budget-based contracts. However, these simulation studies do not rely on the agent-based modeling paradigm where heterogeneous interacting agents are explicitly modeled and where agents' actions are the main trigger of the models' dynamics.

From its very `technical´ core ACE relies on simulation and, thus, has a key feature in common with ``non-ACE´´ simulation-based research in MAR - including also the general pros and cons of simulation-based research as described before.

However, comparing ACE applied to MAR with the ``non-ACE´´ simulation-based MAR we argue that regarding construct validity both approaches are characterized by similar virtues whereas a trade-off between internal and external validity shows up. In particular, with respect to construct validity both approaches require a computational representation which compels the researcher to precisely specify constructs in terms of variables and algorithms. Regarding external validity, ACE typically employs models of agents who show more cognitive limitations (e.g., biases, illusions) than the ones included more traditional models of simulation-based MAR. This contributes to more realistic models of decision-makers in that bounded rationality in their numerous facets can be captured and thus external validity increases. However, this may come at a cost of internal validity compared to non-ACE simulations: as was mentioned before, capturing boundedly rational agents in computational models is among the key challenges of ACE potentially leading to arbitrariness and loss of rigidity. It requires including sub-models of agents' sensing, learning and prediction (see part C of the internet appendix for details); these submodels possibly also have to take dynamic interactions with other agents into account and have to be ``customized´´ for different agents in order to capture agents' heterogeneity. Hence, modelling the multitude of parameters and submodels for a multiplicity of agents may be a virtue but also particular challenge of ACE-based simulations compared to non-ACE-based simulations in MAR.

In sum, we argue that it is predominantly the trade-off between internal and external validity that may shape the researcher's choice between ACE and more traditional simulation approaches in management accounting: *if the research question requires to capture decision-makers and decision-making processes as realistic as possible, ACE appears suitable; in contrast, when aspects of internal validity are paramount, more traditional forms of simulation may be the better choice.*

*Analytical Research compared to ACE*

Guffey and Harp (2017) provide a comprehensive definition of analytical research in MAR: ``Models are developed via applied mathematics and/or logic. The models use analytic devices to hypothesize and test certain interrelationships that predict, explain, or give substance to theory." (p. 94). In previous parts of this paper, we already highlighted various aspects of the two modeling approaches – namely when introducing ACE as compared to more traditional schools of economic thought (see also Table 1).

Regarding construct validity it is helpful to distinguish the intertwined aspects it comprises, (1) to consider if constructs capture the theoretical `logic´ as they should and (2) if they can be measured precisely. Regarding the first aspect, we argue that both analytical modeling and agent-based modeling compel the researcher to precisely specify constructs – though in different ways. Analytical models are constrained by tractability, which, for example, requires specifying a tractable number of variables and functional relations. In agent-based models it is the computational representation which requires precision in terms of, for example, variable declarations and algorithms. Regarding the second aspect of construct validity, i.e., measurement precision of construct, the two modeling approaches do not have to struggle with `noisy´ measurements as in empirical research. However, in ACE, the numerical analysis and presentation of data produced by the simulations is of crucial relevance, which is usually not the case in analytical research. Hence, we argue that ACE requires the researcher to take particular questions of informative and significant metrics into account (see also the design concept ``observation´´ in Table C.1 in the internet appendix C), which is not the case in analytical modeling.

Regarding internal validity analytical modeling allows inferences by means of mathematical derivation and *proofs* (see also Table 1). In contrast, ACE employs numerical experiments where - instead of mathematical rules - algorithms enforce internal validity. However, `numerical experiments´ means that *examples* are provided – even though there may be thousands or millions of them. While this alone suggests that internal validity of ACE is lower than that of analytical modeling, we must keep in mind two further aspects mentioned before, which go in a similar direction: the huge combinatorial space of parameters of simulation models and the fact that the sometimes opaque internal dynamics resulting from agent-based models are overly complicated.

With respect to external validity we argue that ACE is ahead of analytical modeling. Requirements of tractability set constraints not only to the number of variables captured in analytical models but also to the type of behavioral assumptions that can be modeled. In this sense, there have been calls to enrich analytical modeling in order to increase its relevance (e.g., Shields 1997, Hopper and Bui 2016). Key features of ACE which we identified before as making ACE an interesting candidate for MAR (e.g., realistic behavioral assumptions including agents' heterogeneity, rich institutional arrangements and rich contingencies), enabled ACE to be ahead of analytical modeling in respect of external validity. Moreover, the researcher can easily run simulations for a broad parameter space, thus, exploring a multitude of environmental conditions, which serves generalization and external validity.

In sum, we argue that primarily the trade-off between internal and external validity may facilitate the choice between analytical research and ACE. *If questions of internal validity are paramount, analytical research appears more appropriate; in contrast, when the research question requires rich contextualization and realistic behavioral assumptions, ACE is more suitable.*

*Laboratory Experimental Research compared to ACE*

When addressing laboratory experiments in MAR, we refer to research designs in which an experimental task that captures a real-world practice setting is assigned to a population of test persons and where the setting is manipulated. Above all, the researcher studies how manipulations of the setting (treatments) affect the judgments and decisions of test persons (e.g., Bloomfield, Nelson, and Soltes 2016, Smith 2019).

With respect to construct validity, Smith (2019) finds that it typically receives little attention in experimental research – presumably because the method grants tight control to the researcher – and if construct validity is addressed it mainly relates to whether the observable variables are reliable measures for the variables of interest.

The particular strengths of experimental research lie in internal validity: The method elicits effects of independent variables on dependent variables via the manipulation of settings (treatments) and, hence, allows studying causality (Abernethy et al. 1999, Sprinkle 2003, Bloomfield, Nelson, and Soltes 2016). However, it has been argued that there are some particular threats to internal validity in laboratory experiments, most of which relate to the participants over time (Smith 2019, with further references): Apart from challenges related to selection of participants, temporal aspects are regarded as threats to internal validity. These include, for example, maturation (e.g., participants change due to learning across experimental stages), experimental and subject mortality (i.e., in a series of experiments/treatments the experimental environment may have changed or subjects may have moved/left), resentful demoralization (different treatments cause different motivations among participants for subsequent stages of the experiment)

In order to obtain strong causal attributions (internal validity), the experimental settings may be set rather tightly, e.g., abstract experimental tasks for judgment and decision-making which do not provide much current information and context. This in turn may be a threat to external validity in the sense that contextualization is sacrificed (Bloomfield, Nelson, and Soltes 2016).

When comparing laboratory experiments and computational experiments based on ACE, some correspondences turn out to be worth mentioning. Thus, one may argue that treatments in laboratory experiments correspond to variations in those parameter settings in simulation experiments which capture the key constructs of interest. However, while laboratory experiments usually employ few variations of the independent variables (treatments) the usually huge combinatorial space with its numerous feasible parameter settings is a virtue as well as an obstacle of simulation-based research at the same time. One may regard the laboratory experiments' challenges for internal validity with respect to participants as a counterpart to challenges of computational models of humans in ACE. On the other hand, in ACE, the researcher has control over the agents' behavior and, hence, does not have to struggle with temporal aspects as they are related to participants in laboratory experiments. This may be of interest when time plays a relevant role in the research question.

Hence, if the management accounting researcher thinks about choosing between computational experiments in the spirit of ACE and laboratory experiments we recommend considering the particular trade-off between internal validity and external validity in different experimental settings. *Laboratory experiments grant high internal validity for studying the causality among variables via (few) treatments in reduced (abstract) experimental settings of tasks with human participants. In contrast, ACE allows studying a multitude of variations in complex configurations of tasks in rich contextualization with computational models of humans.* However, we believe that the two experimental methods are complements rather than substitutes to each other, which we briefly address below.

*Fieldwork Research compared to ACE*

In this sub-section we intend to elaborate on pros and cons of ACE in comparison to `fieldwork´ research. We employ this term as an umbrella term for `field studies´ and `case studies´ (in a similar vein, e.g., Smith 2019). In field studies, the researcher observes attributes in their natural context without manipulations (treatments) and makes inferences from the observations of a single unit or a few units based on in-depth analysis (e.g., Guffey and Harp 2017, Bloomfield, Nelson, and Soltes 2016, Eisenhardt and Graebner 2007, Eisenhardt 1989). Note that there is a wide variety of forms of fieldwork regarding, for example, involvement of and interventions by the researcher, number of observations and directions of observations (cross-sectional vs. longitudinal). Fieldwork is particularly relevant in the context of grounded theory and interpretative MAR (for further references, see Elharidy, Nicholson, and Scapens 2008, Gurd 2008).

Fieldwork has been argued to allow for detailed observations and, hence, to provide rich contextual insights (Merchant and Van der Stede 2006). In particular, the method allows the researcher to customize observations to those aspects which are of interest for the research question, and this might be relevant in terms of construct validity (Bloomfield, Nelson, and Soltes 2016). However, regarding internal validity, i.e., when inferring from observations, fieldwork is argued to have some weaknesses. The researcher's observations may unintendedly affect what is observed and replicability (whether the same observation would be made by other researchers) is in question (Smith 2019).

Field research employed for theory-building also entails that explanations are inferred from a limited number of observations. On the one hand, this may be seen as an obstacle for internal validity since it hampers considering and potentially eliminating alternative explanations. In a way, this corresponds to the aforementioned limitations of agent-based models: a model in which a certain pattern emerges from the bottom-up shows that an explanation is possible, but neither that this is the only explanation nor that it is a plausible explanation (see the sub-section on ACE's focus on theory-building and the inherent obstacles above).

On the other hand, the limited set of observed settings in fieldwork is regarded as limiting generalizability and, hence, external validity. In this vein, Eisenhardt (1989) recognizes `[t]he risk that the theory describes a very idiosyncratic phenomenon or that the theorist is unable to raise the level of generality of the theory´´ (p. 547). With respect to the number of `observations´, ACE is at the other end of the scale: because hundreds of thousands of simulations could be run for a given configuration of context; and, moreover, the parameters capturing the context can easily be modified and, thus, a multitude of contextual configurations can be studied.

Fieldwork is seen as an inductive approach for theory-building (Eisenhardt 1989, Eisenhardt and Graebner 2007) and, in this sense, has a similar aim as ACE. We conclude that weighing fieldwork vs. ACE primarily means trading off *in-depth analysis of a limited set of particular real-world observations for analysis of large-scale computationally generated data in a multitude of parameterized contexts.*

***Archival Research compared to ACE***

In a broad understanding archival research means that sources, especially data, employed for research are generated from archives (e.g., of historical accounting numbers, databases like Compustat®) which were typically not established for purposes of academic research (Moers 2006, Smith 2019). According to Bloomfield, Nelson, and Soltes (2016), a characteristic of archival research is that the ``researcher delegates recording and possibly structuring of data archives that have been drawn from a practice setting without intervention´´ (p. 365). It is worth mentioning that in MAR a predominant focus of archival studies is put on *theory-testing*, particularly in the field of management compensation (Moers 2006, Lachmann, Trapp, and Trapp 2017).

It has been argued that archival research faces particular challenges in respect to construct validity and internal validity. The former is due to the fact that data employed were not created in respect to the particular research question (see the example of proxies for earnings management in Dechow, Ge, and Schrand 2010) and that it might be difficult or even impossible to reconstruct what aspects have been considered or excluded at the time when the data were recorded (Smith 2019). Hence, as aforementioned, this constitutes a contrast to ACE, which – as based on a simulation model – grants the researcher a high level of control.

Regarding internal validity, it is noteworthy that the researcher's lack of intervention makes it difficult to find causal relationships where only associations among constructs can be observed. However, the lack of the researcher's intervention is regarded as a virtue with respect to construct validity (Lachmann, Trapp, and Trapp 2017) and with respect to strengthening generalization and, hence, external validity (Smith 2019).

Archival studies are often of longitudinal nature, which Lachmann, Trapp, and Trapp (2017) see as a strength with respect to internal validity; moreover, it has been argued that it might be possible to select data in such a way that exogenous shocks which serve as a natural manipulation are covered as well (Bloomfield, Nelson, and Soltes 2016). In respect to external validity, archival research usually benefits from large samples which may relate to various contexts and allow discovering / testing for associations among constructs. Regarding

ACE, it should be recalled that processual and longitudinal perspectives are a key feature of simulation-based research.

Hence, compared to ACE, we argue that the main trade-off occurring is among the researcher's *high level of control in respect to constructs, internal logic and large-scale `artificial´ data on the side of ACE versus external validity based on intervention-free large-scale empirical data, whereby both methods allow for a longitudinal perspective.*

***Survey Research compared to ACE***

In survey research, the researcher records and structures data on dependent variables from respondents without manipulating independent variables in terms of treatments (Bloomfield, Nelson, and Soltes 2016, Guffey and Harp 2017). In management accounting the vast majority of survey-based studies is used for *theory-testing* (Van der Stede, Young, and Chen 2005), whereas the focus of ACE is on *theory-building*.

In 2018, JMAR published a ``Special Interest Forum on Survey Research´´, and in the introductory article Speklé and Widener (2018) state that ``survey research provides researchers with the ability to tap into relatively complex, multi-faceted phenomena as they occur in their natural setting´´ (p. 3). This allows for rich contextualization and generalizability and, hence, high external validity of survey research. Regarding construct validity, Bedford and Speklé (2018) conclude that – in spite of sophisticated measurement techniques – the key question of whether constructs represent what they should may require more attention.

Van der Stede, Young, and Chen (2005) find that in survey research cross-sectional designs predominate by far compared to longitudinal designs. The authors argue that this is due to the particular challenges that longitudinal designs pose in surveys, which are to either repeat surveys over time or to ask respondents in one-time surveys about measurements over time. Both variants suffer from threats like increasing non-response over time in the former case and recall bias in the latter. It has been argued that longitudinal designs provide greater confidence for causal inferences (internal validity) than cross-sectional designs because they more easily establish temporal priority (Pinsonneault and Kraemer 1993).

In contrast, studying processual and longitudinal phenomena is one of the particular strengths of simulation-based research in general (Davis, Eisenhardt, and Bingham 2007) and – as elaborated in the previous sections of this paper – is one of the key features of ACE for MAR. *We conclude that ACE may be particularly more suitable than a survey-based study when the research question requires a longitudinal perspective in various settings and rich contextualization in the researcher's control, which, however, comes at the cost of artificial instead of real-world data.* However, we think that survey-based research and ACE are complements rather than substitutes, not least because of the different foci on *theory-testing* versus *theory-building* in ACE.

**Combining ACE with Other Methods for MAR**

The comparison of ACE with other research methods in MAR shows that in some cases the choice in favor of or against ACE means trading off internal versus external validity; partially, it is an `intra-type of validity´ trade-off. (e.g., human agents in narrow contexts in lab experiments vs. computational models of humans in wide contextualization with ACE). However, in this sub-section we argue that it may be interesting to also consider ACE as a complement rather than a substitute to other methods and, in particular, suggest some paths for combinations of methods. With this, we follow Labro (2015a) when she calls for bridging the divide between different methods in MAR while, at the same time, noticing the particular challenges related to combining different methods (e.g., Malina, Nørreklit, and Selto 2011, Hoque, Covaleski, and Gooneratne 2013).

With a focus on theory-building, we believe that a combination of analytical research and ACE provides interesting research opportunities. ACE may be employed to study how robust results of analytical models relying on `heroic´ behavioral assumptions are when these assumptions are relaxed in an agent-based model. The agent-based version of the standard hidden-action model (Holmström 1979) as outlined in part D of the internet appendix is an example (Leitner and Wall 2020). We believe that incomplete contracts could be another topic in MAR which could benefit from ACE. With incomplete contracts – broadly speaking – a contract does not stipulate the appropriate actions for the contracting parties for each possible future event and, hence, grants the contracting parties discretional action for cooperative or opportunistic behavior (Baiman 1990, Tirole 2009). Incomplete contracting is predominantly studied by analytical methods (e.g., Hart and Moore 1988, Tirole 2009) or by means of experimental research (Malhotra and Murnighan 2002, Christ, Sedatole, and Towry 2012). The key features of ACE described in Section 4 reveal a ``natural fit´´ to key issues of incomplete contracting. For example, the emergence of cooperation – which is a key issue in incomplete contracts – is among the most `traditional´ topics studied by means of agent-based modeling (Axelrod and Hamilton 1981); bounded rationality and cognitive limitations which are a cause for incomplete contracting (e.g., Tirole 2009) are a core element in ACE; environmental turbulence inducing renegotiations among contracting parties can be captured in agent-based models as well as rich institutional arrangements in which incomplete contracts are embedded. Moreover, incomplete contracts and, in a broader sense, relational contracts may enfold and be subject to complex social dynamics: building of (dis-)trust, recognition of reputation and trustworthiness, formation of trust networks or self-reinforcement (e.g., Cook and Gerbasi 2009, Coletti, Sedatole, and Towry 2005). With this, ACE could bridge between analytical research (with the tight behavioral assumptions) and experimental work related to individuals' behavior (derived for particular environmental and institutional contextualization in the experiments).

Regarding laboratory experiments there is a stream of research which stresses the cross-fertilization with ACE (e.g., Duffy 2006, Janssen and Ostrom 2006). For example, data from laboratory experiments can be used to calibrate or test ACE models of individual decision-making or formation of expectations (which refers to bounded rationality as being a key feature in ACE, see the fourth section).

In a similar vein, one may argue that survey research may be combined with ACE (beyond theory-testing). As Speklé and Widener (2018) stress, surveys are ``especially useful for obtaining data on respondents' perceptions, attitudes, and beliefs that drive their behavior´´ (p. 3). Hence, these attributes of the agents' `acting´ in an agent-based model may be based on survey research, including indications as to what may affect these attributes. Opinion dynamics and emergence of social norms are among the key topics studied by ACE and a combination with survey-based research may be particularly promising to study cultural controls in organizations.

We suggest also considering combinations of ACE with fieldwork. For example, the researcher may be interested in finding out whether an eventually idiosyncratic explanation (Eisenhardt 1989) for a certain observation may apply to other contexts, i.e., in exploring the explanation's generalizability. Then, ACE may provide a means for, first, studying whether the observation can be reproduced in a model - i.e., `grown´ in the words of Epstein (1999); if so, the researcher can, second, explore further contexts by variation of parameters in order to figure out under which conditions the phenomenon observed in the field emerges.

In a similar vein, ACE may be combined with archival or survey-based research. Findings derived from large-scale empirical research may let the researcher seek to explain them in terms of `growing them bottom-up´. The second example in part D of the internet appendix introduces a model providing some explanations for the emergence of tight coordination which refers to survey-based research (Davila 2005). However, in many cases it is the other way round: via agent-based simulation certain hypotheses are derived which await testing by means of appropriate empirical research methods.

# CONCLUSION

The aim of this paper was to provide an overview of opportunities and difficulties that ACE may entail for research in management accounting. While ACE has been relatively often adopted in other domains of managerial science, it is a fairly new paradigm for MAR.

Our main contribution is the introduction of a framework that researchers in management accounting may employ when considering ACE as a paradigm for their particular research endeavor. The framework builds on the two interrelated paradigmatic aspects incorporated in ACE: first, a set of theoretical assumptions whose core is the bounded rationality of economic actors with its consequences for actors' heterogeneity and their way of problem-solving; second, ACE relies on agent-based modeling which differs from other prominent modeling paradigms in economics. From its particular properties, we propose five key features that ACE may have for MAR, namely

- realistic behavioral assumptions about agents' rationality,

- rich institutional arrangements and rich interactions,

- rich contingencies,

- processual perspective: learning and emergence, and

- bridging the micro-macro-divide.

We discuss how these key features could be valuable for studying recent topics in management accounting. We further compare ACE to other methods in MAR with particular focus on the various trade-offs regarding internal and external validity and seek to provide some suggestions for combining ACE with other methods in MAR.

We hope that researchers will find the proposed framework helpful for further research, in particular when considering whether ACE could be an appropriate approach for their research endeavor. However, given the low dissemination of ACE among researchers in management accounting so far, we guess that ACE's further `fate´ in the domain depends on several aspects. One of the key questions is, of course, whether researchers (and journals) are interested in adopting ACE as a research paradigm, and, consequently, in fathoming its value in MAR. This is a prerequisite not only for exploration of opportunities of ACE in MAR but also for mitigating its weaknesses for the domain. Take, for example, the high flexibility incorporated in agent-based modeling which, as we tried to outline, is a blessing as well as a curse. Higher adoption among researchers in management accounting, however, may result in the evolution of certain standards and protocols in its application to topics of MAR. Standards, in turn, enable fostering acceptance and transparency in communication of research results. We hope that this paper may contribute to such a promising perspective of ACE in MAR.

# APPENDIX

to

F. Wall; S. Leitner: Agent-based Computational Economics in Management Accounting Research: Opportunities and Difficulties, *Journal of Management Accounting Research.* doi: 10.2308/jmar-19-073

**Part A: Brief Introduction to Cellular Automata and NK Fitness Landscapes**

This part of the appendix briefly introduces the two structured simulation approaches which are most widely applied in agent-based models in managerial science, i.e., cellular automata and NK fitness landscapes.

*A.1 Cellular Automata*

*Cellular automata* (Wolfram 1986, for overviews Dooley 2002, Fioretti 2013, Chen 2016, pp. 93) consist of a grid where each agent "resides" in a cell and, hence, the lattice reflects a spatial distribution of the agents. The distribution can be regarded as spatial in the literal sense of the word, or could also capture, for example, the cultural distance among agents. The cells can take various states (most simply just the states "0" or "1"). The state $s_{j,t}$ of cell $j$ at time $t$ depends on its own state in the previous period $t$-1 and the previous states of the neighboring cells, for example, cells $s_{j-1, t-1}$ and $s_{j+1, t-1}$. This shows the two key features of cellular automata for ABM (Walker and Dooley 1999). First, the impact of agents on each other depends on the distance between them, i.e. the closer the neighborhood, the greater the influence on each other. Second, the state of a cell (agent) is specified in rules representing the agent's behavior. For example, one rule might be "if the sum of left and right neighbors is two, change to a 1", and another "if the sum of left and right neighbors is lower than 2, change to a 0", where 1 could mean ``adopting Balanced Scorecard´´ and 0 ``not adopting Balanced Scorecard´´ and each cell could represent a business unit. Hence, starting from an initial configuration which may be randomly chosen, the states of the cells in the grid change over time according to the specified rules. From the macro perspective of the grid, i.e. the system to be investigated, the researcher is interested in whether certain patterns occur from the interaction processes and, if so, how. Hence, agent-based models employing cellular automata often study the emergence of macro-level patterns resulting from local interactions like competition, diffusion or segregation in a set of agents.

*A.2 NK Fitness Landscapes*

In contrast, *NK fitness landscapes* provide a `built-in´ structure which constitutes a particularly interesting approach when processes of search, optimization and decision-making are in the center of research questions (Davis, Eisenhardt, and Bingham 2007). NK fitness landscapes were originally developed in evolutionary biology (Kauffman 1993, Kauffman and Levin 1987) to study how effectively and how fast biological systems adapt to reach an optimal point. Levinthal (1997) introduced NK-fitness landscapes to managerial science where they have been widely used in the meantime (for overviews, Ganco and Hoetker 2009, Wall 2016). The term NK fitness relates to the number $N$ of attributes (e.g. activities, decisions) and the level $K$ of interactions among these attributes. Each attribute $i$ can take two states $d_i \in \{0,1\}, i = 1,...,N$ and, hence, the overall configuration **d** is given by an $N$-dimensional binary vector $\mathbf{d} = (d_1,...,d_i,...,d_N)$. The state $d_i$ of attribute $i$ contributes with $C_i$ to the overall fitness $V(\mathbf{d})$ of configuration **d**, where `fitness´ stands for the particular objective that is relevant in the particular context of interest (e.g., firm performance). However, depending on the interactions among attributes, $C_i$ is not only affected by a single attribute $d_i$, but also by the state of $K$ other attributes $d_{j, j \neq i}$. In the case of $K = 0$, the fitness landscape is single-peaked. If $K$ is raised to the maximum, i.e. $K = N - 1$, altering one single state $d_j$ affects the fitness contributions of all other attributes and, usually, this leads to highly rugged fitness landscapes with numerous local maxima for $V(\mathbf{d})$ (Altenberg 1997, Rivkin and Siggelkow 2007).

The explicit modeling of interactions among attributes may explain the value of NK fitness landscapes for research in managerial science: via parameter *K,* the approach allows studying systems with variable complexity in terms of interdependencies among subsystems (e.g., among decisions to make in an organization) with respect to overall fitness. Hence, NK fitness landscapes provide a basis to capture a task environment and to conveniently shape its complexity. Within the environment, the effectiveness of strategies of search for superior solutions and institutional arrangements in which search takes place (e.g., incentive schemes, lateral coordination) are the center of interest.

**Part B: On the Implementation of Agent-Based Models**

The purpose of this part of the appendix is to provide the interested reader with some further information related to implementations of agent-based models. For the novice in ABM, we believe that three aspects may be particularly relevant: first, getting a `feel´ for agent-based models thanks to examples of running models; second, gaining some hands-on experience; third, getting further information regarding implementation and software platforms.

*B.1. Examples of Agent-Based Models `at Work´*

For ABM, the software platform NetLogo developed by Uri Wilensky (with further references in Wilensky and Rand 2015) is among the most widely used; moreover, textbooks in ABM rely on this platform (e.g., Railsback and Grimm 2019). The platform is freely available on the NetLogo website (http://ccl.northwestern.edu/netlogo/, last access: August 26, 2020) and it can be either downloaded or run online in a browser. The NetLogo platform includes a large library of sample models. In most sample models the user can easily modify key parameters (e.g., number of agents, observation time) at the user interface and, thus, can study the effects of parameters on the outcome. For the very beginning, the model of wolves and sheep as an example of predator-prey models could be a good start:

http://ccl.northwestern.edu/netlogo/models/WolfSheepPredation (last access: August 26, 2020). The NetLogo library also comprises various models from the domain of economics and financial markets which can be found under the term "Social Science" at:

http://ccl.northwestern.edu/netlogo/models/ (last access: August 26, 2020).

*B.2. Hands-on Experience in Implementation*

The implementation of an agent-based model requires translating the conceptual model which may be specified verbally and / or partially in mathematical notation into a piece of software that can be executed by a computer. Hence, implementation often means programming / coding in a programming language. However, for a novice to ABM even without particular coding experiences it may be of interest to implement a first simple model.

For this purpose, we recommend employing a familiar spreadsheet software: with its structure of cells and interactions among cells spreadsheet tools allow to easily represent a cellular automaton as introduced in this appendix. We refer the reader to Hand (2005) which is also available online (https://www.economicsnetwork.ac.uk/cheer/ch17/hand.htm, last access: August 26, 2020):

Hand (2005) takes the reader through the implementation of a model that captures the diffusion of a product innovation on a spreadsheet in two variants – a deterministic and a stochastic variant of a cellular automata. By varying the number of early adopters, their proximity and, in the stochastic version, their followers' probability of adoption, the reader can observe some properties of the diffusion processes.

## B.3 Implementation and Software Platforms

The researcher interested in employing ABM can find various resources online.[1] However, new platforms are introduced, software platforms are frequently updated, model libraries extended and new software comparisons and reviews published. In view of recent updates and a dynamic research community, we recommend the reader the respective websites of Leigh Tesfatsion – one of the pioneers of ACE.[2]

While her site on ACE (http://www2.econ.iastate.edu/tesfatsi/ace.htm, last access: August 26, 2020) provides a general overview, she also maintains a site on ``General

---

[1] The reader may for example have a look at the wikipedia site https://en.wikipedia.org/wiki/Comparison_of_agent-based_modeling_software, last access August 26, 2020), which compares agent-based platforms and provides further links.

[2] With this recommendation we follow the American Economic Association (https://www.aeaweb.org/rfe/showCat.php?cat_id=88, last access: August 26, 2020) and the Society for Computational Economics (http://comp-econ.org/#, last access: August 26, 2020).

Software and Toolkits´´ (http://www2.econ.iastate.edu/tesfatsi/acecode.htm, last access: August 26, 2020). Here, the reader first finds a large body of material which comprises reviews and comparisons of software platforms, links to certain software components like random-number generators and tutorials. Moreover, the website provides a comprehensive list of software and toolkits that are used by ACE researchers with links to the respective sites of platforms.

Those readers who are particularly interested in the implementation of design concepts as introduced in our design guidance in Appendix C (esp. Table C.1), may refer to Railsback and Grimm (2019). In their book, the authors provide advice on how to implement most of the design concepts with NetLogo and present related models.

**Part C: Design Guidance for Employing ACE in Management Accounting Research**

In this part of the appendix, we put particular emphasis on choices concerning key elements of the *design* of an agent-based model. Further choices to be made related to the *implementation* of a model (i.e., software platform, programming language) were discussed in part A of the internet appendix.

*C.1 Some Remarks on the ODD Protocol and the ODD+D Protocol*

In order to make agent-based modeling more rigorous and comprehensive, the concept of pattern-oriented modeling has been introduced – meaning that a model is composed and communicated based on its characteristic and essential underlying structures and processes (Grimm et al. 2005). In this vein, from its introduction in 2005 and with several updates, the ODD-protocol has become a widely accepted structure for developing and describing agent-based models (Grimm and Railsback 2005, Railsback and Grimm 2019, Grimm et al. 2020). The ODD protocol consists of the three building blocks, `overview´, `design concepts´ and `details´, the three initial letters of which make up the term:

- It is a key aspect for gaining an `*overview*´ of a model to first state the questions the model is intended to provide answers for. Hence, the *purpose* of the model and the principle components to be captured in the model should be clarified, as well as the main state variables should be introduced. In an ABM this includes, foremost, clarifying which agents are captured in the model, i.e., the low-level entities like managerial decision-makers, accountants, auditors, and whether they are embedded in a hierarchical context or not. To gain a principled overview of an ABM it is also recommended to identify the environmental and individual processes which are captured in the model. These could be, for example, firm growth, production of goods, processes of decision-making, reporting.

- The `*design concepts*´ capture the modeler's design choices which are related to typical features of agent-based models as elaborated in the field of agent-based modeling. Hence, design concepts represent features of agent-based models which are particularly characteristic for this modeling approach. These include, for example, emergence (i.e., system's outcome at the macro level resulting from agents' behavior at the micro level) or adaptation (e.g., agents' reactions to changes in their environment).

- Beyond the `design concepts´, various `*details*´ are to be specified in an ABM like how variables are initialized and which input of data is required to run a simulation.

The ODD protocol has been introduced and extensively used for agent-based models in ecology (e.g., Polhill et al. 2008). However, with respect to agent-based models capturing human decision-making, some complementary elements were added, especially regarding the `design concepts´ part of the ODD protocol – resulting in the so-called ODD+D protocol (Müller et al. 2013). In the subsequent sub-section, we introduce the design concepts of the ODD+D protocol with particular focus on design choices to be made by the researcher when employing agent-based modeling. Building on the ODD+D protocol corresponds to the plea of Labro (2015) to (re-)shift the focus of MAR towards decision-making (see also Section 2).

*C.2 Design choices in agent-based models for MAR based on decision concepts*

In the following, we introduce the ten design concepts captured in the ODD+D protocol (Müller et al. 2013) as they capture essential design choices which the researcher has to make when applying ABM to research questions where human decision-making agents are involved.[3] In particular, we seek to map the ten standard design concepts of the ODD+D protocol to the framework of key features of ACE for MAR as elaborated in section 4. Table C.1 summarizes the subsequent considerations.

---

[3] Note that not all design concepts are necessarily relevant to each agent-based model.

*1. Theoretical and empirical background*

The purpose of this design concept is to clarify the basic principles in terms of theoretical and empirical foundations the agent-based model particularly reflects. For example, regarding the modeling of decision-makers in an organization, different variants of bounded rationality (Selten 2002) could be employed and captured in computational models of decision-making (Johnson and Busemeyer 2010, Busemeyer and Johnson 2004). This also includes specifying the behavioral assumptions related to the agents of the model and establishing how these assumptions are empirically founded, e.g., by laboratory experiments. In a similar vein, the principled understanding of the institutional arrangements decision-makers are faced with, should be determined – be it, for example, the information-processing view (Galbraith 1974), the neo-institutional view (for an overview Alvesson and Spicer 2019) or evolutionary theories (Dosi and Marengo 2007) of organizations. Correspondingly, with respect to capturing contingency factors in agent-based models in MAR, the particular stream of contingency view on organizations and management control is to be clarified (Van de Ven, Ganco, and Hinings 2013, Otley 2016). While emergence at some system's level is at the very core of ACE, a specification of what is regarded as micro and macro is required – particularly since this is likely to entail further theoretical consequences. For example, as Molloy, Ployhart, and Wright (2011) show, regarding the market as ``macro´´ differs from considering society as ``macro´´ and, hence, different theoretical backgrounds (economics vs. sociology) may be appropriate for the interpretation of results. Hence, from these considerations, we believe that for the design of an agent-based model in MAR specifications regarding the theoretical and empirical backgrounds is relevant in respect to each of the five key features as elaborated before (cells 1.a to 1.e in Table C.1).

The designer of a model has two different possibilities to incorporate basic principles in the model's design. First, the principles can be implemented in the model ``by design´´. For example, decision-makers may be modeled to make their choices according to rank-dependent utility theory (Johnson and Busemeyer 2010). Second, the basic principles could be the centre of the research and, hence, the model may capture alternative basic principles in order to compare the evolving effects against each other. For example, a model may capture rank-dependent utility of decision-makers and, alternatively, also configural weight utility (Birnbaum et al. 1992) and compare how these alternative substantiations of prospect theory affect the outcome at the system's level.

*2. Individual decision-making*

Agent-based models employed in MAR are likely to comprise decision-making agents like single managers, boards, accountants or auditors. With respect to agents' decision-making, when designing the model the researcher has to specify the following aspects:

- The decision-making *subjects* (agents) and *objects* of decision-making. Objects could be prices or quantities of products to be sold, reward systems to be implemented or certification of the financial statement. Interestingly, agents may also decide on other agents; for example, a manager may decide which subordinate manager to hire.

- The various decisions captured in the model may they be nested due to *multiple levels* or hierarchical structure. For example, choices made at a higher level may affect decisions at lower levels in terms of setting constraints.

- The *objectives* which decision-makers pursue when making their decisions. In MAR, these objectives may be driven by incentives, performance measures, social or organizational norms. The modeler also has to specify how the objectives are pursued: for example, agents may be optimizers according to an explicit objective or they may be satisfied when certain aspiration levels are met.

- The *rules of how decision-making* agents make their choices. Based on the behavioral assumptions made, the researcher may capture more or less sophisticated decision-making rules of agents, e.g., full optimization or the use of some heuristic.

- How agents *adapt to environmental changes*. What characterizes the environment depends on the particular model; examples are budget constraints or social norms. The environment in which an agent acts may be subject to changes –be they driven by exogenous or endogenous factors. When adapting to a changing environment, the agent may make use of previous experiences or expectations for the future, which requires the modeler to endow decision-makers with memorizing or forecasting capabilities, respectively. Moreover, changes in the environment may also enter agents' decision-making in terms of uncertainty which could be captured in the decision-making rules as mentioned above or in agents' learning (see below).

Hence, with ACE allowing to capture behavioral assumptions which go beyond traditional economic thought (see Table 1 in section 3), modeling of agents' decision-making provides a broad space of design choices regarding objectives and rules of decision-making (cell 2.a in Table C.1). The *subject-object combination* of decision-making, *objectives* shaped by incentive schemes as well as *multi-level structures* of decision-making in the model relate to the modeler's design choices regarding capturing (rich) institutional settings as differentiation, delegation of decision competencies and hierarchical structures in the model (cell 2.b in Table C.1). Specifying agents' *adaptation to changing environments* contributes to capturing rich contingencies and processual phenomena in a model (cells 2.c and 2.d in Table C.1).

*3. Learning*

Decision-making agents may change their decision-rules based on learning. For example, an agent may learn which decision-rules contribute better / best to desired outcomes based on experience (i.e., a type of reinforcement learning). Another way of learning could be that of imitating other agents' behavior, particularly if these agents are successful with respect to achievement of their objectives. Hence, if the model is intended to capture learning-based dynamics in terms of agents' decision-rules, the modeler has to specify the *type of learning* to be employed in the model. There is a broad range of learning modes in ACE differing, for example, with respect to whether learning takes place unconsciously, is based on routines or on beliefs (for further classifications of learning see Brenner 2006).

The researcher may not only intend to capture a single agent's learning, because even *collectives* of agents may learn. Organizational learning could be modeled as a collective of agents who learn by transferring experience into routines which guide behavior of agents within that organization (Levitt and March 1988). For modeling organizational learning, the modeler has to decide how experience is to be shared among agents, how they are to agree on inferences and transfer the inferences into routines for further decision-making (e.g., Marengo 1992). With respect to MAR, the emergence of collusive behavior within a collective of managers is a relevant topic (e.g., Evans III et al. 2016). If this is to be studied by means of an agent-based model, the researcher has to determine how agents exchange information, learn to interact, cooperate or collude.

The design concept of `learning´ as outlined before could be employed as one element to capture (more) realistic agents – who are learning how to make choices (cell 3.a in Table C.1). Moreover, when collective learning is modeled this may be applied to studying the emergence of institutional arrangements (routines) via organizational learning (cell 3.b in Table C.1). Learning is a means of inducing processual phenomena and the type of learning employed may shape the emerging processes (cell 3.d in Table C.1).

*4. Sensing*

Agents behave in response to other agents or to their environments. The researcher needs to decide how agents sense – or get knowledge about – their own state variables, those of other agents and those of the environment. In particular, the researcher has to make several design choices which relate, at least, to the content and costs of sensing, communication processes and channels, precision, and temporal structure:

- *Content*: for each agent, it has to be determined which variables capturing their own state, the state of other agents and of the environment an agent can sense. With respect to MAR, this includes, for example, whether an agent representing the head of a business unit knows certain accounting numbers related to their own business unit, and knows which choices the other units' heads have made or how demand for products has changed.

- *Costs of sensing*: `sensing´ may come along with costs – in terms of cognitive effort or costs for gathering information. The researcher specifies whether agents should consider costs of sensing. If so, sensing may even evolve endogenously resulting from an agent's considerations on whether it is worthwhile to gain further information or not – which is a topic in the context of incentive schemes (e.g., Lewis and Sappington 1997), and more broadly in the context of boundedly rational decision-makers following Simon's (1955) satisficing approach (Caplin, Dean, and Martin 2011).

- *Communication processes and channels*: in order to allow the agents in a model to `receive´ signals for sensing, the modeler has to decide on the respective communication channels. A first choice is whether agents simply `know´ about the information; if so, communication is not modeled explicitly, but communication and reception of signals is implicitly regarded as `happening´. However, there may be reasons to explicitly capture communication processes and channels in a model – for example, in order to depict the formal organizational structure as well as social networks existing in a firm.

- *Precision*: Sensing of state variables may be erroneous when different types of errors are of interest. For example, there may be imprecisely ``measured´´ accounting numbers that a decision-maker employs (e.g., Labro and Vanhoucke 2007). Imprecise sensing may also result from failures in communication due to errors on a sender's side (Wall 2019b) or in the communication channel. Other types of errors may result from misperceptions on the receiver's side with systematic misperceptions referring to cognitive biases (Gerber and Green 1999, Pronin 2007, Pohl 2017) and, thus, contribute to capturing realistic behavioral assumptions. The modeler, hence, may not only have to decide on the level of imprecision in sensing but also on its causes, which also leads to the question of whether erroneous sensing is regarded an exogenous or endogenous variable in a model.

- *Temporal structure of sensing*: the modeler has to decide on when an agent gets to know about states or changes of states. For example, it might be reasonable to assume some time delay: head of department A realizes choices of department head B one period after they were made, which induces further decisions on A's side which B becomes aware of with one period delay, leading to further decisions by B and so forth. In a similar vein, decision-makers may get to know about changes in their broader environment (e.g., technological changes, new competitors) with time delay. Hence, such a temporal structure of sensing may induce a chain of mutual and time-delayed adjustments, which contributes to depicting processual phenomena.

Based on these considerations, we argue that `sensing´ as a design concept contributes to the implementation of key features of ACE for MAR: capturing `sensing´ in a model means noticing that receiving information is an issue in its own right, which may be prone to errors and incurring costs, both of different causes and types. This contributes to reflecting realistic behavioral assumptions in the model (cell 4.a in Table C.1). As far as a model comprises communication channels this is a means to depict rich institutional arrangements, e.g.,

Appendix - 14

organizational structures as well as social networks in the model (cells 4.b and 4.c in Table C.1). Since sensing of new information may require time, reflecting time-delays in a model is more than a further piece of more realistic behavioral assumptions (cell 4.a in Table C.1) – it is also a means for the researcher to capture a processual perspective (cell 4.d in Table C.1) in the model.

*5. Prediction*

Agents may employ predictions about future states of the environment, about other agents' choices and about the consequences of their own decisions. Hence, the researcher has to specify in which way agents acting in the model estimate future states. A first design choice is whether agents' predictions are modeled explicitly or, instead, implied only "tacitly" in simple rules (Railsback and Grimm 2019). An example for the latter case would be to let each decision-making agent conjecture that the fellow agents will not undertake any further actions in the next period – which may however lead to surprises afterwards. In case that the modeler chooses to explicitly model agents' predictions, several aspects are to be specified. As such, an *agent's internal model* for predictions has to be defined. For example, an agent may form expectations based on exponential smoothing of past observations. This directs to the design choice about the *data* an agent uses for prediction and the agents' *memory*. Another aspect to be specified is whether and, if so, how agents' predictions of the consequences of the decisions are *erroneous*. Prediction may be subject to agents' learning; agents may learn which mode of prediction works better.

The aforementioned design choices contribute to implement agents with realistic behavioral assumptions due to erroneous predictions, simple rules of forecasting or limited memory (cell 5.a in Table C.1). Moreover, agents' predictions may not only concern results of their own decisions but also future states of an (external) environment, e.g., demand for products, technological change. Hence, agents' predictions serve as a means to take a contingent view (cell 5.c. in Table C.1). Depending on the agents' internal models prediction also serves to induce path dependency and, thus, a processual perspective (cell 5.d in Table C.1).

*6. Interaction*

Local interactions among agents are among the core elements of agent-based modeling and, in a broad sense of meaning, the term ``interactions´´ refers to agents affecting each other's state. The term ``local´´ relates to a distance in a spatial or figurative meaning (e.g., colleagues in a department). The designer has to make design choices with respect to various aspects of interactions.

- *Network and type of interactions*, i.e., which agents may interact in which way with each other. Agents may interact directly, for example, when one agent sells a good to another agent. In contrast, indirect interactions are mediated by the environment, be it within an organization or in an external environment. For example, two departments may compete for the financial budget provided by the company or for highly qualified workers on the job market.

- *Representation of real interaction mechanisms*: In MAR direct interactions could be reporting to a superior, informing colleagues or rewarding subordinates. Indirect interactions could result from participating in an auction, decisions in transfer pricing, or compensation offers in highly competitive labor markets.

- *Conditions under which interactions become active:* for example, in each period each department informs the headquarter about the choices made in that period without any exceptions. Alternatively, this report depends on whether, or not, the overall performance achieved in that period is below a certain threshold. Hence, due to some stochasticity in the model (see below) not all interactions among agents necessarily have to be ``active´´ in any period or simulation run. For example, when predictions are subject to some errors (see above) it may be that a manager-agent erroneously assumes that a certain threshold of demand will be exceeded and, thus, starts hiring workers on the job market that another department – based on its predictions – also wants to hire.

- *Communication processes and channels*: Here, we refer to the aforementioned design concept of `sensing´.

- *Temporal structure of interactions*: for example, in one time-step of the simulation clock the superior may fix the value base for rewarding the subordinates and credit the reward to each subordinate's bonus bank. The payment from the bank is deferred for some time steps, which may be of interest when combined with behavioral aspects like (im-)patience and costs of retention of managers.

- *Coordination network*: Particularly with respect to agent-based models related to decision-making agents, the ODD+D protocol proposes specifying whether a coordination network is captured in the model (Müller et al. 2013), which relates to whether a ``(de)centralised or group-based coordination structure of the agents exists´´ (p. 43). Following well-known concepts of ``management control systems´´ (Merchant and Van der Stede 2017, Kruis, Speklé, and Widener 2016, for a review see Strauß and Zecher 2013), aligning the behavior within an organization, and particularly that of decision-makers with respect to overall objectives is at the very core of these systems. Hence, with respect to the coordination network facet of the ``interaction´´ design concept, we argue that a modeler in MAR should specify which particular controls (i.e., results, action, personnel and cultural controls following Merchant and Van der Stede 2017) or subsystems (beliefs, boundary, diagnostic control or interactive control system according to Simons 1994) the agent-based model is intended to capture. For example, the modeler may wish to specify a combination of sequential planning (action control) where, at the same time, some results controls including a reward structure are applied and some organizational norms as cultural controls are set. The coordination network may be exogenously defined or may be an emergent property of the model.

Design choices related to the interactions contribute to implementing rich institutional arrangements as captured in Management Control Systems (cell 6.b in Table C.1). In particular, indirect interactions, i.e., interactions mediated by the environment, are a means to represent rich contingencies (cell 6.c in Table C.1). Since interactions may have non-trivial temporal structures, they provide a way to induce processual effects (cell 6.d in Table C.1). Finally, the direct and/or indirect interactions among agents are a main driver to induce emerging properties at a system's level and, thus, bridging the micro-macro divide (cell 6.e in Table C.1).

*7. Collectives*

The researcher may wish to group individual agents into collectives. Collectives may be social groups or the members of a business unit. The agents who are assembled in a collective share some common properties (e.g., same cohort of hiring date). By employing collectives the researcher can introduce one or even more intermediate levels in a model – i.e., between the individual agents and the overall system. For capturing collectives, the researcher has to make some design decisions.

First, the researcher has to specify whether the collectives are defined by the modeler or whether they emerge in the course of the simulations. For example, the modeler may define a certain structure of departments with each department comprising several agents; the modeler may want to study how a certain structure of departments emerges from adaptive and learning processes of agents. Second, one individual may belong to several collectives. For example, an agent representing a management accountant may belong to the collective of colleagues in the respective business unit, to a network of all management accountants across business units in the firm and, the same time, might be part of an advocacy group for employees with a specific cultural background. Third, how will the collective affect the behavior of its members and vice versa. For example, the aforementioned network of management accountants may agree on employing certain standards reporting procedures, which then affects how its members communicate accounting information to superior managers in their business units. Vice versa, the success story of a novel tool one accountant mentions in an accountants' meeting may encourage others to follow and thus induce some dissemination in the organization. Hence, employing collectives in a model calls for the researcher specifying further interactions.

Collectives are a means to capture rich institutional arrangements and contingencies in a model (cells 7.b and 7.c in Table C.1). Moreover, collectives are a way of mediating between micro-level and macro-level and, hence, of bridging the micro-macro-divide (cells 7.e in Table C.1).

*8. Heterogeneity*

Heterogeneity of agents is among the core properties that the agent-based modeling paradigm offers to the researcher and, with respect to MAR, particularly heterogeneity related to agents' decision-making may be of relevance (Labro 2015). Against this background, a first question that the researcher has to decide on is whether the agents `acting´ in a model should be heterogeneous. This requires the researcher to specify which parameters and/or processes show differences across agents. With respect to management accounting these could be, for example, cognitive capabilities, understanding of management accounting techniques or objective functions. Moreover, agents may suffer from different decision-making biases, tend to employ different decision-making rules or be differently responsive when learning how to decide in turbulent environments (see above design concepts `individual decision-making´ and `learning´).

Hence, ACE abandoning the idea of the so-called representative agent (Kirman 1992) as familiar in more mainstream thought of economics (Table 1) in favor of heterogeneous agents marks a turn towards more realistic behavioral assumptions on economic decision-makers. This is why we argue that heterogeneity as a design concept in agent-based models primarily contributes to this feature of ACE (cell 8.a in Table C.1).

*9. Stochasticity*

Agent-based models usually employ some stochastic processes, which requires that the modeler makes some appropriate design choices. These choices include specifying *which components and processes* of the model (including the initialization) are subject to randomness and for which *reason*. When capturing stochastic elements in a model, the modeler has to select an appropriate *statistical distribution* for the random numbers which capture stochasticity. Empirically determined probabilities could provide a valuable basis to reproduce observed behavior (Müller et al. 2013).

With respect to *reasons* for including stochasticity, assumptions related to limited cognitive capabilities and bounded rationality appear particularly relevant for MAR. As mentioned before, a decision-maker's errors in decision-making, sensing and prediction (see the respective design elements 2, 3, 4 and 5 above) may be based on concepts of bounded rationality including various types of imprecisions and biases as elaborated in related research. In this sense, stochasticity is a means to capture realistic behavioral assumptions in MAR (cell 9.a in Table C.1).

The researcher may want to capture some variability in the model because variability is regarded as being important, but without including the causes for variability in the model, e.g., price fluctuations may be an important component whereas the causes for price volatility are not. In this sense, stochasticity could also contribute to capturing rich contingencies in a model (cell 9.c in Table C.1).

Moreover, the *temporal structure* related to stochasticity may be employed to strengthen a processual perspective in an agent-based model for MAR in several ways (cell 9.d in Table C.1). First, stochasticity contributes to inducing sequences of actions that agents may take. For example, instead of agents that are capable of overseeing the entire space of feasible options as in more mainstream schools of economic thought, agent-based models usually discover the solution space stepwise. Which options are discovered in the search processes might be guided by the type of search (e.g., in the neighborhood) but also affected by some randomness. Hence, together with limited knowledge of the solution space agents *proceed* with searching for superior options. Second, stochasticity itself could be subject to emergence. For example, the clues (e.g., metrics provided by management accounting) which decision-makers in a model use for their choices may improve over time in terms of lower noise due to learning on how to generate the clues (e.g., by evolving improvements in management accounting). Third, the researcher may wish to let the system undergo some external shocks to induce turbulence from time to time and thus study the system's speed of adjustment to changes.

*10. Observation*

This design concept is related to the outputs provided by the model. The `observed´ data should correspond to the research question which the model is intended to address. Studying the emergence of properties at the system`s level from the actions and interactions of individuals is a core feature of agent-based modeling. In particular, the researcher has to specify which outputs the simulations should produce in order to observe the internal dynamics of the model and the emerging behavior at the system's level. Hence, outputs from the model should allow analyzing and understanding emergence and, thus, bridge the micro-macro-divide (cell 10.e in Table C.1). This includes not only the observation of emerging properties at the macro-level (e.g., performance achievements of the entire organization) but also requires observing processes and – if subject to emergence – properties of agents and interactions at the micro-level.

Moreover, the researcher has to decide which data at which time in the simulation are to be collected. For example, does it suffice to record which accounting-based metrics turned out to be useful for decision-making in the final period of the simulation runs, or should the processes of learning on the metrics in an organization also be recorded. Recording time series of data may be part of adopting a processual perspective and allows tracing dynamics and understanding emerging properties (cell 10.d in Table C.1).

**(<<<< insert Table C.1 >>>>)**

# TABLE C.1

# Design Guidance According to the ODD+D Protocol for ABM in MAR

| Design concept | Relevant design choices for MAR | Key features of ACE for Management Accounting Research | | | | |
|---|---|---|---|---|---|---|
| | | a. Realistic behavioral assumptions | b. Rich institutional arrangements and rich interactions | c. Rich contingencies | d. Processual perspective | e. Bridging the micro-macro divide |
| 1. Theoretical and empirical background | - general concepts underlying the model, particularly with respect to<br>  - agents' behavior and principle decision models<br>  - organizational design,<br>  - contingency view,<br>  - conceptualization of the overall system (macro)<br>- implementation of general concepts "by design" or as subjects of emergence | X | X | X | X | X |
| 2. Individual decision-making | - subjects and objects of decision-making ("which agent should decide on what objects")<br>- levels / nesting of decisions<br>- objectives decision-makers pursue and social norms they seek to comply with<br>- decision-making rules<br>- adaptation processes to changes in the environment | X | X | X | X | |
| 3. Learning | - type of how single agents learn about their decision-rules<br>- learning of collectives, organizational learning | X | X | | X | |
| 4. Sensing | - content, i.e., which states of its own, other agents and the environment each agent senses<br>- costs of sensing<br>- communication processes and channels<br>- precision, e.g., noise or biases<br>- temporal structure, e.g., time delays | X | X | X | X | |
| 5. Prediction | - explicit or `tacit´ type of prediction<br>- agents' internal models including data and memory employed for prediction<br>- precision, e.g., noise or biases<br>- agents' learning about how to predict | X | | X | X | |
| 6. Interaction | - network and type of interactions<br>- real mechanisms to be captured<br>- conditions for activation<br>- communication channels<br>- temporal structure<br>- coordination network | | X | X | X | X |
| 7. Collectives | - properties and structure of collectives<br>- emergence or pre-definition of collectives<br>- relation between collective's and individuals' behavior | | X | X | | X |
| 8. Heterogeneity | - properties and processes subject to heterogeneity across agents and for decision-making with particular respect to design concepts 2, 3, 4 and 5 | X | | | | |
| 9. Stochasticity | - components/processes subject to stochasticity including statistical distributions<br>- reasons for stochasticity<br>- temporal structure of stochasticity | X | | X | X | |
| 10. Observation | - outputs (data) to be observed on system's properties (`macro´) and on internal dynamics<br>- temporal structure of observed data (data points vs. time series) | | | | X | X |

**Part D: Examples of Studies in Management Accounting inspired by ACE**

This part of the appendix outlines two examples of studies in MAR in the spirit of ACE. Table D.1 summarizes the studies according to the framework of key features proposed in section 4 (see also Table 2).

*D.1 Example 1: Agentization of the Standard Hidden-Action Model*

A large proportion of research in management accounting is based on agency theory (Gomez-Mejia, Berrone, and Franco-Santos 2014, Bosse and Phillips 2016) and core issue is to identify the most efficient contract between the principal and the agent which incorporates a set of assumptions that mainly cover the principal's and the agent's behavior and the availability of information (Müller 1995, Cuevas-Rodríguez, Gomez-Mejia, and Wiseman 2012). On the one hand, these assumptions allow for deriving optimal contracts in rigorous closed-form modeling; on the other hand, they may limit the theory's predictive validity (Eisenhardt 1989, Lambright 2008, Roberts and Ng 2012). The positive agency literature proposes relaxing these specific assumptions in order to increase the theory's explanatory power (Cuevas-Rodríguez, Gomez-Mejia, and Wiseman 2012). ACE, and in particular the `agentization´-approach (Guerrero and Axtell 2011, Leitner and Behrens 2015) allows doing so by transferring closed-form agency models into agent-based models.

The following outlines key-features of an agent-based `version´ of the standard hidden-action model, and presents a snapshot of results of the study presented in Leitner and Wall (2020). Particular focus is put on the first feature of the framework presented in section 4, i.e., realistic behavioral assumptions about agent's rationality; a summary is provided in Table D.1.

In the standard hidden-action model (Holmström 1979) both the principal and the agent are assumed to be rather `gifted´: they possess information about the entire space from which the agent can select the effort, have information about the distribution of the environment, and know the exact production function. In the standard hidden-action model, the principal, in addition, is fully informed about the agent's characteristics, such as the utility function, reservation utility, and productivity. There is information symmetry except for the effort exerted by the agent and the realization of the environment, where the agent can observe the latter ex-post (Lambert 2001). The resulting incentive scheme can, thus, only be based on performance outcomes. This results in a strong focus on the measurability of the outcomes (Hesford et al. 2007, Lambert 2001), and, thus, on the decision-influencing role of information (Demski and Feltham 1976). Regarding decision-facilitating information, the standard hidden-action model simply assumes that both principal and agent have all the information[4] necessary to make `optimal´ decisions in one shot, or, are capable to process all information available to them. This, however, is not necessarily in line with human capabilities (Eisenhardt 1989, Hendry 2002).

The agent-based model introduced in Leitner and Wall (2020) shifts the attention from the decision-influencing to the decision-facilitating role of by taking limited information-processing capabilities of the contracting parties into account. In particular, the principal's and the agent's information about the environment and the action space are limited. The contracting parties are, however, endowed with three additional capabilities: First, as they have limited information about the environment, they are capable to learn about the environment over time. Second, they can decide to either exploit the area of the action space which they already know or to explore the remaining, so far unknown, fraction of the action space. Third, they are endowed with memory in which they can store their learnings.

---

[4] It is worth mentioning that this refers to the standard hidden-action model while there are numerous principal-agent models where decision-facilitating information is not available entirely, has to be learned over time or is asymmetrically distributed among principal and agent (e.g., Baiman and Sivaramakrishnan 1991, Hemmer and Labro 2019, for an overview Holmbert 2001).

The agent-based model of the standard hidden-action problem also includes a shift in the perspective on the model's outcome: as the principal and the agent have only limited information about the environment and the action space but can learn the `missing´ pieces of information, they cannot find the optimal solution immediately. Rather, they have to search for the optimal solution or, in other words, they can only optimize with limited information. Their state of information is, however, stochastically affected by their capabilities to learn and the availability of different search strategies, which allows for observing the emergence of an incentive scheme (including its convergence to the `optimal´ incentive scheme).

The resulting model is limited in its institutional arrangements (as there is only one principal and one agent) but allows investigating interactions among different types and states of information. In addition, a large number of contingencies, like different characterizations of the environment (stable vs. dynamic) and different cognitive abilities can be captured for the contracting parties. The latter may be either different memory capacities or different fractions of the action space which can be overseen at once (Leitner and Wall 2020). Regarding the micro-macro divide, the model allows for establishing a link between individual characteristics at the micro-level (e.g., learning mechanisms, states of information) and performance management at the macro-level.

The results presented in Leitner and Wall (2020) suggest, amongst others, that turbulent environments exert a certain pressure to be innovative on managers, resulting in almost immediate performance increases, which, however, flatten very soon, so that no further performance increases can be expected after only a few periods. In stable environments, on the contrary, this pressure is missing. Performance, therefore, steadily increases in small steps. The results also indicate that the choice of search strategy (exploration or exploitation) affects performances significantly at the macro-level only if decision-makers are well-informed about the environment. Compared to the optimal contract proposed by the standard hidden-action model (Holmström 1979), in the agent-based version *on average* a contract emerges which induces an agent's effort that leads to a performance below the optimal one. In other words, on average the emergent contract is less efficient than the optimal contract. An in-depth analysis of emerging contracts during single simulations runs, however, indicates that some contracts emerge which provide the agent with incentives to make even *more* effort, which is no longer optimal for him. The latter observation is mainly driven by the extent of environmental turbulence: The higher the turbulence the more pronounced the effect.

### *D.2 Example 2: Emergence of Boundary Systems in Growing Firms*

The second example puts particular focus on emergence of MCS from coevolution of changing contingencies and learning-based change, i.e., the fourth key feature of ACE in MAR (see section 4, Table 2). Motivation, a sketch of the model and key results of the study (Wall 2019a) are outlined in the following with a summary in the far right column of Table D.1.

A common – though rarely explicitly indicated – assumption in the domain of management control is that the MCS of an organization is systematically designed `in advance´. However, high complexity of an organization's overall task, management techniques and behaviors, combined with environmental turbulence and uncertainty, suggest that the MCS may evolve due to learning and adaptation to altered contingencies over time (Chenhall 2003, Fisher 1995, Van de Ven, Ganco, and Hinings 2013). According to the seminal empirical study of Davila (2005), firm size has a positive impact on the overall level of use of MCS. A key argument is that when companies grow, informal controls become too costly or ineffective and, hence, more formal controls are employed, as captured in MCS.

Against this background, Wall (2019a) employs an agent-based model in the tradition of ACE to study, in particular, which type of boundary system – following Simons' (1994) Levers of Control-framework – emerges in the course of firm growth. According to Simons (1994) boundary systems are intended ``to set limits on opportunity-seeking behavior´´ (p.7) and to ``delineate the acceptable domain of activity for behavior of organizational participants´´ (p. 39). However, in growing firms, the MCS is subject to a certain tension, i.e., balancing the necessity to search for novel solutions with mechanisms to deal with increasing intra-organizational complexity (Bisbe and Otley 2004, Widener 2007, Bedford 2015, Kruis, Speklé, and Widener 2016).

For this reason, apart from (growing) firm size, the agent-based model employed considers *task complexity* and the *organizations' search strategy* to find new solutions for their tasks as contingencies.[5] In the model (building on NK fitness landscapes, see section A.1 of this appendix), the *search strategy* as being enforced by the boundary system can be of an exploitative, explorative or ambidextrous nature. Within this frame, that part of the boundary system which prescribes how final choices are coordinated among decision-makers is subject to learning-based emergence in the course of growth. The feasible coordination modes captured in the model range from hierarchical coordination over lateral coordination via sequential planning to, in fact, providing no coordination among units' decisions (for details see also Wall 2018).

Regarding the processual structure, the model comprises three (nested) processes with different time horizons: (1) In the short term, subordinate decision-makers search for superior performance to their particular sub-problems while, depending on the coordination mode effective at the particular point in time, the headquarters eventually may intervene. (2) In the mid-term, the mode of coordination that the headquarters employ for coordination among the subordinate units' choices may be subject to change based on reinforcement learning on the headquarters' side. (3) In the long term, the organizations grow, meaning that the number of task elements and, accordingly, the number of units increase.

---

[5] The agent-based model comprises a growing task environment, task decomposition and delegation of sub-tasks to units.

The results obtained via simulation suggest that the boundary systems emerging in the course of firm growth differ remarkably depending on the level of task complexity which the growing organizations face and depending on the search strategy enforced. For example, when organizations face highly complex increasing tasks – e.g., with complexity corresponding to reciprocal interdependencies according to Thompson (1967) – tight coordination via hierarchy predominantly emerges in the course of growth; moreover, this predominance turns out to be more pronounced when more change is enforced by the search strategy. This suggests an interesting interaction within the boundary system in growing firms: more flexibility and change enforced via search strategy (e.g., exploration) is counterbalanced by tighter coordination (e.g., via hierarchy). This corresponds to the aforementioned particular tension captured in MCS in growing firms, i.e., allowing and enforcing novelty and setting behavioral constraints via coordination. Regarding the design of the boundary system within MCS, the results suggest considering the internal fit between search strategy enforced and ways of coordination prescribed. This, in particular, also relates to the third key feature of ACE in MAR as elaborated in section 4 (see Table 2).

**(<<<< insert Table D.1 >>>>)**

# TABLE D.1

# Overview of the Illustrative Studies According to the Key Features of ACE in MAR

| Key Features of ACE | Example 1: Agentization of the Standard Hidden-Action Model | Example 2: Emergence of Boundary Systems in Growing Firms |
|---|---|---|
| 1) Realistic behavioral assumptions about agents' rationality | - principal and agent have limited information about the environment<br>- principal and agent cannot oversee the entire space of feasible effort levels but only know a subset of levels | - units cannot oversee entire solution space `at once´ but search it stepwise<br>- agents (units and headquarter) cannot evaluate newly discovered solutions perfectly<br>- units cannot foresee their fellow units' choices |
| 2) Rich institutional arrangements and rich interactions | - delegation relationship with one principal and one agent | - hierarchical organizations with growing number of units;<br>- decomposition of the organization's overall task into sub-tasks;<br>- incentive scheme rewarding performance of units;<br>- behavioral rules regarding coordination among units (subject to learning) |
| 3) Rich contingencies | - different characterizations of the environment<br>- different memory capacities for the principal and the agent<br>- different capabilities to oversee the space of feasible effort levels | - organizational size (subject to growth)<br>- level of task complexity: (1) perfectly decomposable; (2) non-decomposable, based on NK fitness landscapes<br>- search strategy employed by units: (1) exploitative; (2) explorative; (3) ambidextrous |
| 4) Processual perspective: learning and emergence | - mechanisms to learn about the environment over time<br>- exploration and exploitation as search strategies | - short-term: adaptive search for higher levels of performance by steepest ascent hill-climbing<br>- mid-term: learning on the coordination mode captured in the boundary system based on reinforcement<br>- long-term: growing firm size in terms of growing number of task elements and number of units |
| 5) Bridging the micro-macro-divide | - establishes a link between individual characteristics of the contracting parties at the micro-level and the performance management system at the macro-level | - establishes a link between individual choices, coordination among various subordinate decision-makers at a meso-level, and adaptation of the management control system at the macro-level |